# Depth-Resolved Evolution of Buried Polar Topologies in a $PbTiO_3/SrTiO_3$ Superlattice

Xinxin Hu[1,*], Penghan Lu[2], Noa Varela-Dominguez[3], Anthony Edgeton[4], Chang Beom-Eom[4], Francisco Rivadulla[3], José Santiso[1], Yingzhuo Lun[1,6], Zhihua Sun[5], Rafal Dunin-Borkowski[2], Gustau Catalan[1,7], Jordi Arbiol[1,7,*]

[1]Catalan Institute of Nanoscience and Nanotechnology - ICN2 (CSIC & BIST), Barcelona, Catalonia 08193, Spain

[2]Ernst Ruska-Centre for Microscopy and Spectroscopy with Electrons, Forschungszentrum Jülich, Jülich, Germany

[3] CiQUS, Centro Singular de Investigacion en Quimica Bioloxica e Materiais Moleculares, Departamento de Quimica-Fisica, Universidade de Santiago de Compostela, Santiago de Compostela 15782, Spain

[4]Department of Materials Science and Engineering, University of Wisconsin-Madison, Madison, WI 53706, USA

[5]Chinese Academy of Sciences Fujian Institute of Research on the Structure of Matter, Fuzhou 350108, China

[6] School of Aerospace Engineering, Beijing Institute of Technology, Beijing 10081, China

[7]Institució Catalana de Recerca i Estudis Avançats (ICREA), Barcelona 08010, Catalonia

*Corresponding authors email: xinxin.hu@icn2.cat, arbiol@icrea.cat

**ABSTRACT:** Polar topologies in complex oxides gives rise to a rich spectrum of emergent functionalities and are fundamentally governed by three-dimensional (3D) atomic structures. However, direct experimental determination of buried 3D polar configurations remains a longstanding challenge because conventional (scanning) transmission electron microscopy ((S)TEM) provides primarily projected structural information with limited depth sensitivity. Here, we combine depth-sectioning low-angle annular dark-field (LAADF) STEM, high-angle annular dark-field (HAADF) STEM, and multislice electron ptychography (MEP) to directly visualize the depth-dependent atomic structure and polarization topology in a $[(PTO)_{15}/(STO)_{15}]_{15}$ superlattice. Depth-sectioning STEM reveals pronounced focal-depth-dependent contrast variations and apparent splitting of Pb atomic columns, indicating significant structural heterogeneity along the beam direction. MEP reconstruction simultaneously resolves the Pb, Ti, and O sublattices with nanometer-scale depth resolution, enabling quantitative mapping of atomic displacements throughout the reconstructed volume. The resulting three-dimensional atomic model reveals substantial depth-dependent displacements of Pb, Ti, and O atoms and a corresponding evolution of the polarization topology. Vortex-like polarization structures are observed near the specimen surfaces but become strongly suppressed within the interior, where distinct polarization configurations emerge. These findings demonstrate that polarization patterns observed in conventional projection images can arise from the superposition of multiple depth-dependent polar states and may therefore obscure the underlying three-dimensional polarization texture. Our findings establish a direct experimental link between local atomic displacements and depth-dependent polarization topology, opening new opportunities for investigating and engineering buried functional states in complex oxide nanostructures.

**Introduction:**

Recent studies have revealed that ferroelectric polarization can organize into a rich variety of topological structures, including flux-closure domains[1, 2], vortices[3], polar skyrmions[4, 5], labyrinthine domains[6, 7], merons[8], hopfions[9] and other complex polarization textures[10-12]. Owing to discontinuities in polarization and charge distributions, these non-uniform polar configurations can exhibit emergent functionalities that are fundamentally distinct from those of bulk ferroelectric domains[13, 14], offering exciting opportunities for next-generation electronic and information-storage devices[15, 16].

Despite their scientific and technological importance, direct experimental determination of the three-dimensional (3D) atomic structures of these polar topologies remains highly challenging. Recent advances in scanning transmission electron microscopy (STEM) have enabled atomic-scale characterization of polarization-related properties, including atomic displacements[17,18], strain fields[19, 20], elemental distributions[21], and charge ordering[22]. Typical approach for 3D imaging using conventional depth-sectioning STEM requires the acquisition of focal-series datasets by sequentially shifting the probe focus through the specimen thickness[23-26]. This approach is dose-inefficient and often suffers from significant inaccuracies in crystalline materials because of strong electron-beam channeling effects[27]. Multislice electron ptychography (MEP) has recently emerged as a powerful alternative[28-31]. By combining four-dimensional (4D) STEM datasets with advanced phase-retrieval algorithms, MEP enables quantitative 3D atomic structure reconstruction with deep sub-ångström lateral resolution[32], nanometer-scale depth resolution[33, 34], and enhanced sensitivity to subtle atomic displacements[35, 36], thereby overcoming the projection limitations of conventional STEM.

In this work, we initially employed depth-sectioning annular dark-field scanning transmission electron microscopy (ADF-STEM) to investigate a $[(PTO)_{15}/(STO)_{15}]_{15}$ (PTO/STO) superlattice. For PTO/STO superlattices with this periodicity, complex three-dimensional polar topologies have previously been reported[6,37]. As the probe focus is scanned through the specimen thickness, pronounced contrast variations are observed in low-angle ADF (LAADF) STEM images, while apparent splitting of Pb atomic columns emerges in high-angle ADF (HAADF) STEM images. These observations provided evidence for depth-dependent atomic displacements and structural variations along the beam direction. Motivated by these findings, we subsequently applied multislice electron ptychography (MEP) to quantitatively reconstruct the three-dimensional atomic structure of the PTO layers with simultaneous visualization of both cation and oxygen sublattices. The reconstruction revealed substantial depth-dependent displacements of Ti, Pb, and O atomic columns, which remain obscured in conventional projection images. By quantitatively mapping the displacement of Pb relative to the O sublattice throughout the reconstructed volume, we demonstrate that the polarization topology evolves significantly along the depth direction. Vortex-like and sinusoidal polarization structures are observed near the specimen surfaces, whereas the vortex order becomes substantially weakened or disappears within the interior region. These results reveal that the projected polarization topology arises from the superposition of distinct depth-dependent polarization configurations, providing direct experimental evidence that conventional projection measurements can blur the true three-dimensional polar structure of ferroelectric superlattices. More broadly, our work establishes a general framework for resolving hidden polarization textures in complex oxides and highlights the importance of three-dimensional atomic-scale characterization for understanding and engineering topology-driven functionalities in materials.

## Results

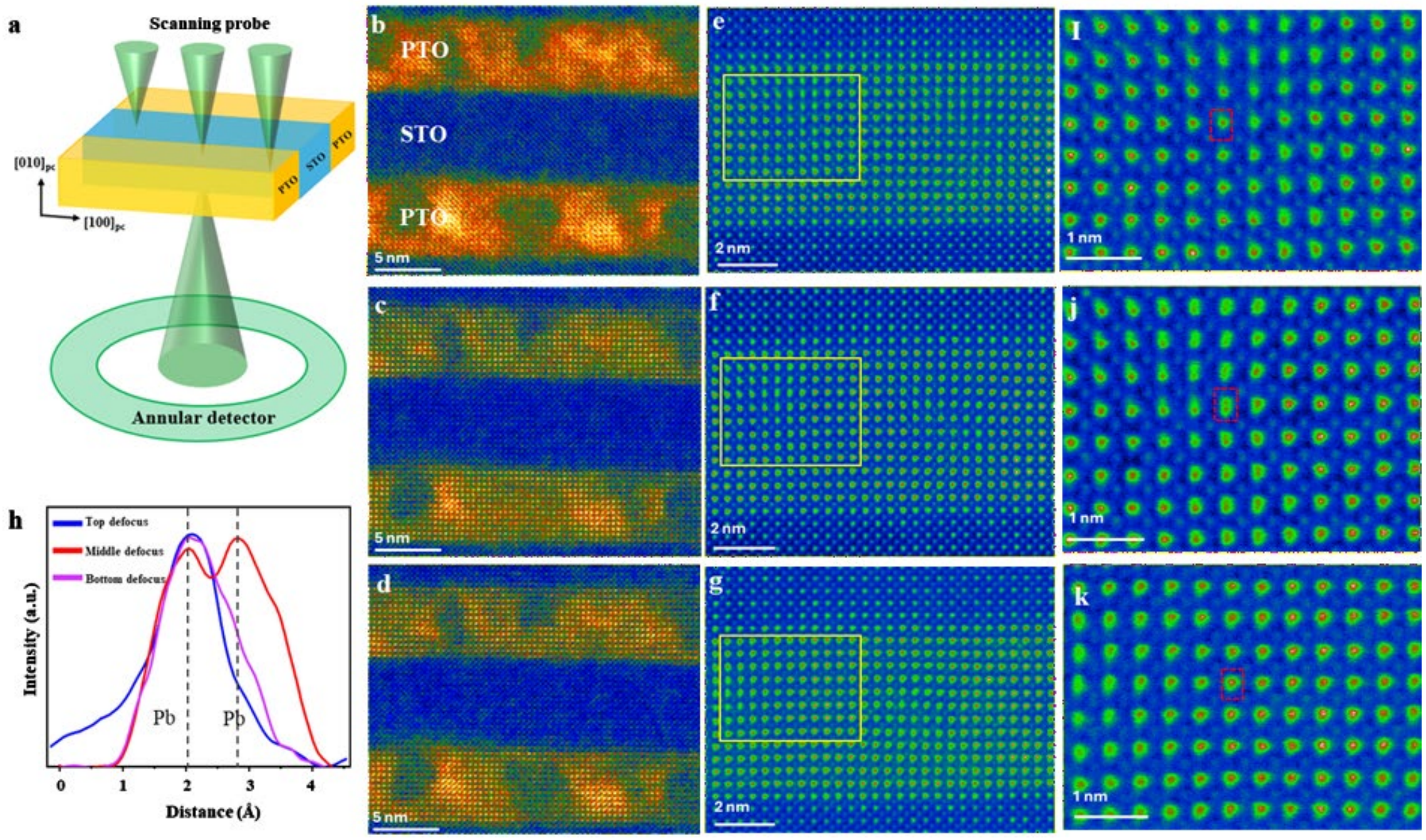

**Figure 1. Depth-sectioning ADF-STEM imaging of the PTO/STO superlattice.** **a,** Schematic illustration of the depth-sectioning ADF-STEM experiment. A convergent electron probe is sequentially focused at the top surface, middle, and bottom surface of the specimen lamella while scanning the same region. **b–d,** LAADF STEM images acquired with the probe focused on the top surface (**b**), middle (**c**), and bottom surface (**d**) of the specimen, revealing depth-dependent contrast variations. **e–g,** Corresponding HAADF STEM images recorded at the three focal planes, with a PTO layer in the central part of the image. **h,** Intensity map extracted from the region outlined by the red rectangle in **i-k**, highlighting the evolution of atomic-column intensity as a function of focal depth. **i–k,** Magnified views of the selected region from **e–g**, showing an apparent splitting, or lateral displacement, of the Pb atomic columns when the probe is focused near the middle of the specimen.

Figure 1 presents the depth-resolved low-angle annular dark-field (LAADF) and high-angle annular dark-field (HAADF) STEM characterization of the PTO/STO superlattice. To assess the capability of depth-sectioning ADF-STEM for detecting buried atomic displacements, multislice simulations of LAADF-STEM and HAADF-STEM (Figure S1) were first performed using a PTO model containing displaced Ti and Pb atoms located approximately 2.7 nm and 5 nm beneath the top surface, respectively (Figure S1). LAADF-STEM is highly sensitive to lattice distortions, strain fields, dislocations, and other crystallographic defects[38, 39]. The simulations show that LAADF-STEM provides enhanced contrast for both displaced Ti and Pb atoms compared with HAADF-STEM. In contrast, HAADF-STEM offers improved depth localization of the displaced Pb atom but exhibits significantly reduced sensitivity to the displaced Ti atom because of its lower scattering intensity. These results indicate that LAADF-STEM and HAADF-STEM provide complementary structural information and motivate their combined use for depth-resolved characterization of the PTO/STO superlattice. A schematic of the depth-resolved STEM experiment is shown in Figure 1a. During acquisition, the electron probe was sequentially focused through the specimen thickness with a defocus step of 3 nm, enabling imaging at different focal depths. Representative depth-sectioning LAADF-STEM images acquired at different focal depths are shown in Figures 1b–d and S3. The bluish contrast is associated with the STO layers, while the orangish contrast is associated with the PTO layers. Pronounced contrast variations are observed within the PTO layers as the probe focus changes, whereas the STO layers exhibit minimal contrast evolution. This distinct depth-dependent behavior suggests the presence of lattice distortions or strain variations within the PTO layers. Energy dispersive X-Ray (EDX) spectroscopy confirms the high compositional quality of the superlattice and the well-defined spatial separation of Pb, Sr, Ti, and O species (Figure S4), indicating that the observed LAADF STEM contrast variations are unlikely to originate from compositional fluctuations. To obtain more directly interpretable atomic-scale structural information, depth-sectioning HAADF STEM imaging was subsequently performed. Representative images acquired with the probe focused near the top surface, middle region, and bottom surface of the specimen lamella in a selected region of the PTO/STO superlattice are shown in Figures 1e–g, with corresponding magnified views centered on one of the PTO layers presented in Figures 1i–k. When the probe is focused near either the top or bottom surface, well-defined Pb atomic columns are clearly resolved. In contrast, when the probe is focused near the middle of the specimen, the Pb columns exhibit an apparent splitting (Figures 1f, j). This observation provides direct evidence of depth-dependent atomic displacements and indicates that the projected Pb columns contain contributions from laterally displaced atomic positions at different depths within the specimen. Furthermore, a well-defined polar vortex is observed in the middle-focused HAADF STEM image, whereas predominantly sinusoidal polarization patterns[40] are observed in the top- and bottom-focused HAADF STEM images (Figure S5). Although HAADF STEM offers strong sensitivity to heavy elements such as Pb and Ti, the large differences in scattering cross-sections among Pb, Ti, and O make the simultaneous visualization of all atomic species challenging[41, 42], as also confirmed by the simulations. In addition, HAADF STEM images are strongly influenced by electron-beam channeling effects, which further complicate direct structural interpretation. Alternative imaging modes, including annular bright field (ABF) STEM and differential phase contrast (DPC) STEM, provide enhanced sensitivity to lighter elements[43, 44]; however, they remain fundamentally constrained by projection effects and limited depth resolution. These limitations motivate the use of multislice electron ptychography (MEP), which enables quantitative three-dimensional structural reconstruction with simultaneous visualization of both the cation and oxygen sublattices.

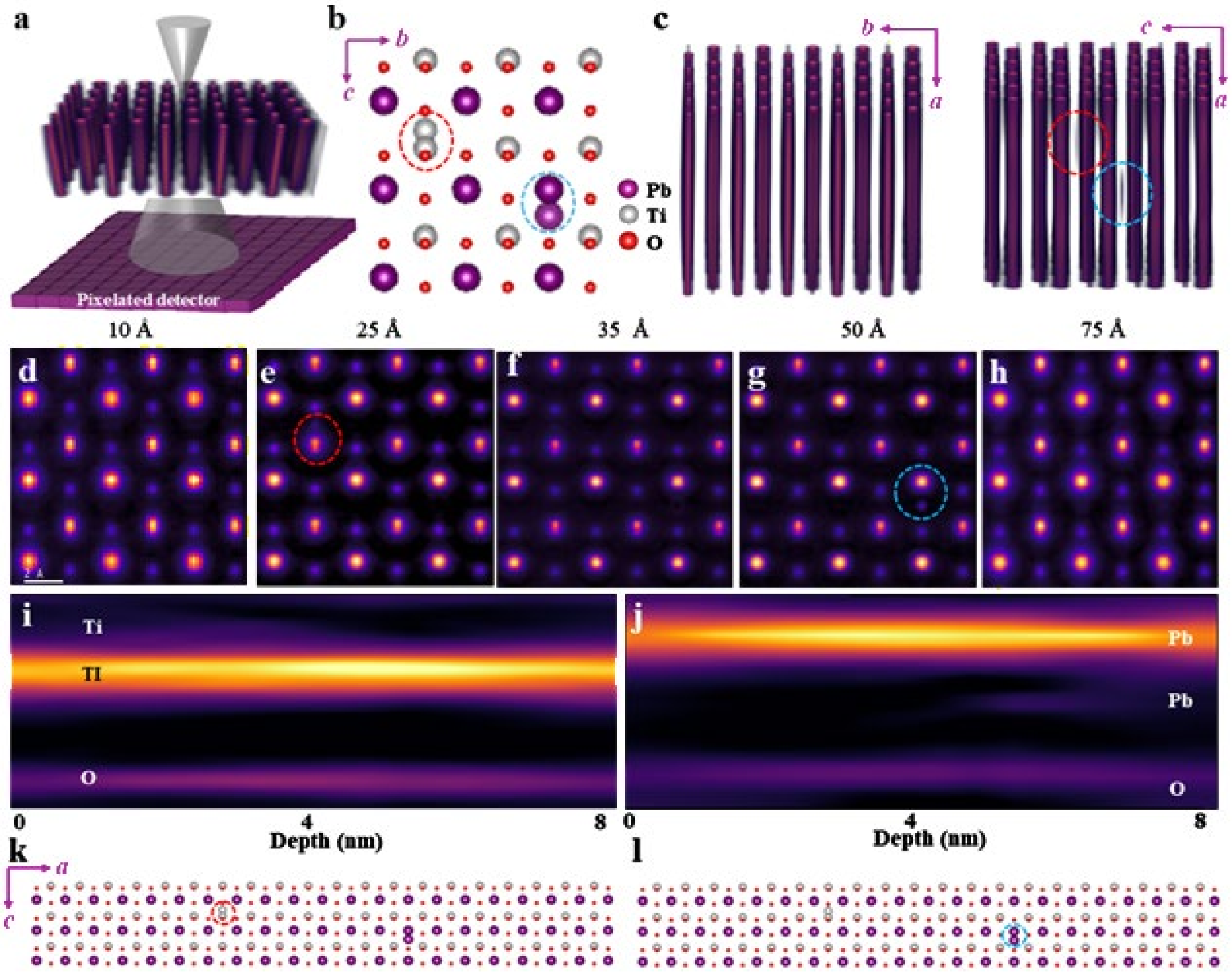

**Figure 2. Simulated multislice ptychographic reconstruction of $PbTiO_3$ with displaced Pb and Ti atoms.** **a,** Schematic illustration of the 4D-STEM ptychography experiment. An over-focused convergent electron probe is scanned across the specimen in real space, while a pixelated detector records the corresponding convergent-beam diffraction patterns in reciprocal space. **b,** Structural model of the $PbTiO_3$ (PTO) sample used for the multislice ptychography simulations viewed along *a*-axis **c,** Three-dimensional phase reconstruction of the PTO sample viewed from different directions, highlighting the displaced Ti and Pb atoms (red and blue dashed circle, respectively). **d–h,** Selected phase-contrast slices extracted at different depths from the reconstructed volume, showing the emergence of the displaced Ti atoms in **e** and the displaced Pb atoms in **g**. **i-j,** Depth-dependent phase-intensity profiles measured along the Ti–O direction marked by the red dash-dotted line in **e** and along the Pb–O direction marked by the blue dash-dotted line in **g**. **k-l,** Structural models of the PTO sample viewed along the *b***-**axis, highlighting the displaced Ti (**k**) and Pb (**l**) atoms, respectively. Red and blue dashed circles indicate the regions with displaced Ti and Pb atoms, respectively.

To evaluate the capability of MEP for resolving buried atomic displacements, simulations were first conducted using the same PTO model (Figure S1). The corresponding 4D-STEM acquisition geometry is illustrated in Figure 2a. Diffraction patterns recorded using a pixelated detector were subsequently used for MEP reconstruction. The resulting phase-contrast volume (Figure 2c) clearly resolves both displaced Ti and Pb atoms, highlighted by the blue and red dashed circles, respectively. Notably, the reconstructed slices (Figure S6) reveal pronounced depth localization of these displaced atomic species. As illustrated in Figures 2d–h, the displaced Ti atoms become distinctly visible at a depth of approximately 25 Å, whereas no discernible signal from the displaced Pb atoms is observed within this slice. At greater depths (~50 Å), the Ti-related displacement contrast progressively diminishes, coinciding with emergence of the displaced Pb atoms. The Pb signal subsequently reaches maximum visibility before vanishing at depths of approximately 75 Å. These observations provide direct evidence of the depth-resolving capability of MEP and its ability to discriminate between atomic displacements occurring at different axial positions within the specimen.

To quantitatively evaluate the achievable depth resolution, regions containing displaced Ti and Pb atoms were selected, and the evolution of their reconstructed intensities was tracked throughout the volumetric dataset. The depth profiles (Figures 2i,j and Figure S7) of the simulated results demonstrate that both displaced Ti and Pb atoms are confined within an axial extent of approximately 2.5 nm. Importantly, these experimentally measured profiles exhibit excellent agreement with the corresponding multislice simulations (Figures 2k and 2l), confirming the reliability of the reconstruction. These results demonstrate that MEP provides nanometer-scale depth resolution while simultaneously resolving both the cation and oxygen sublattices, making it a particularly powerful technique for investigating three-dimensional polarization structures in complex oxide materials.

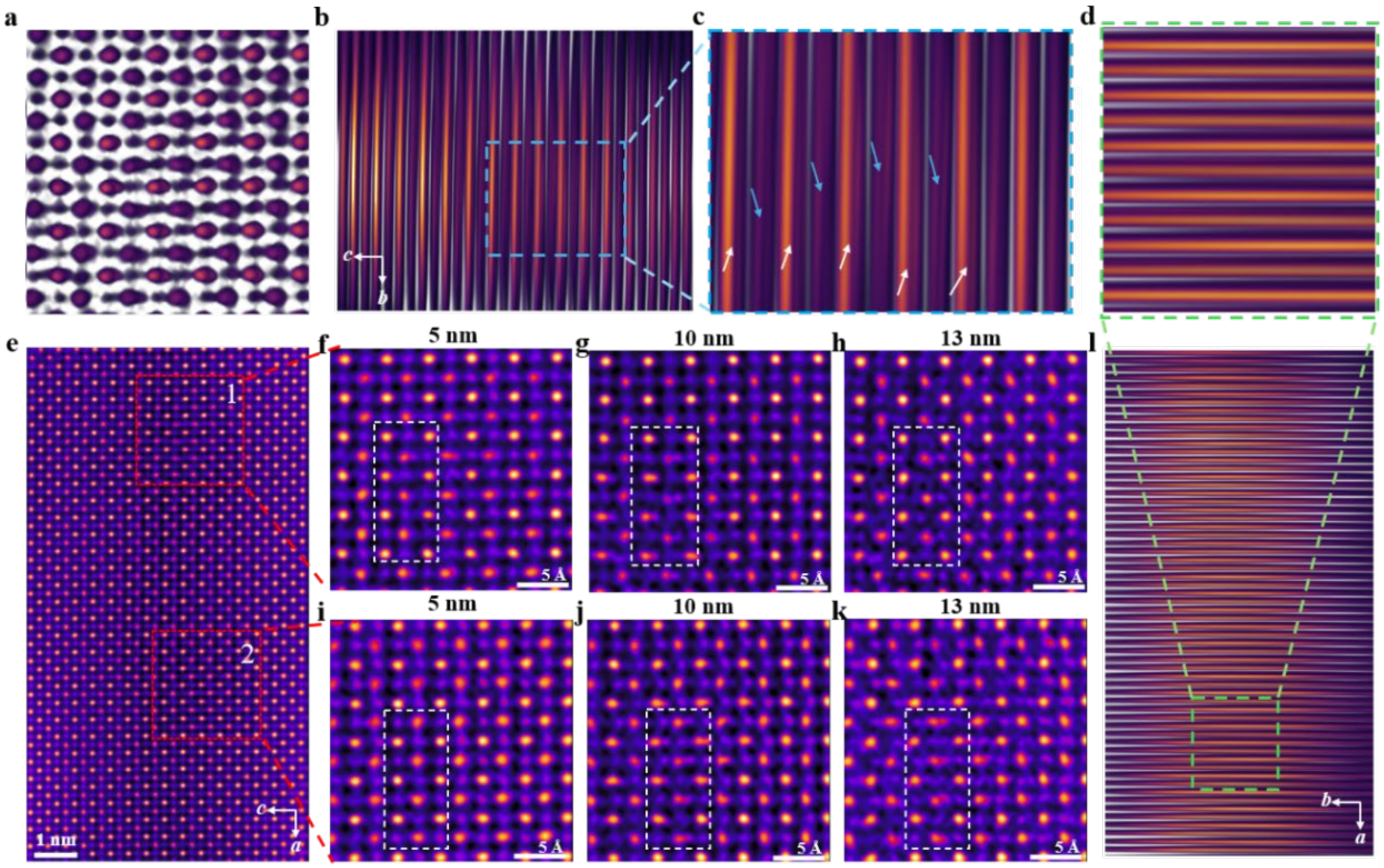

**Figure 3 Experimental visualization of depth-dependent structural variations in the PTO layer of the PTO/STO superlattice.** (a) Summed phase-contrast image of region 1. (b) Reconstructed three-dimensional phase-contrast volume viewed along the *a*-axis and (c) the corresponding enlarged view. (d) Enlarged view of the region highlighted in (l). (e) Summed phase-contrast image highlighting regions 1 and 2 selected for detailed analysis. (f–h) Reconstructed phase-contrast slices extracted at depths of 5, 10, and 13 nm from region 1, respectively. (i–k) Corresponding reconstructed slices extracted at the same depths from region 2. (l) Reconstructed three-dimensional phase-contrast volume viewed along the *c*-axis. The depth-resolved slices reveal pronounced variations in atomic-column configurations along the depth direction that are not evident in the summed projection image.

Having established the depth-resolving capability of MEP, we next applied this approach to a PTO layer within the PTO/STO superlattice. The thickness of the region selected for MEP reconstruction was determined to be approximately 16 nm by combining and electron energy loss spectroscopy (EELS) measurements and Position-Averaged Convergent Beam Electron Diffraction (PACBED) simulations (Figure S8). The resulting three-dimensional atomic reconstruction and the complete set of reconstructed slices are presented in Figure S7.

Figure 3a displays a top-view projection of region 1 obtained from the depth-summed reconstruction shown in Figure 3e. An apparent splitting of the atomic-column contrast is observed, suggesting the presence of depth-dependent lattice distortions. The corresponding *a*-axis projection of the reconstructed volume (Figure 3b), together with the enlarged view, reveals pronounced splitting and broadening of both Pb and Ti atomic columns along *c*-axis, directly indicating structural variations along the electron-beam direction. The depth-dependent nature of these distortions becomes evident upon inspection of individual reconstruction slices.

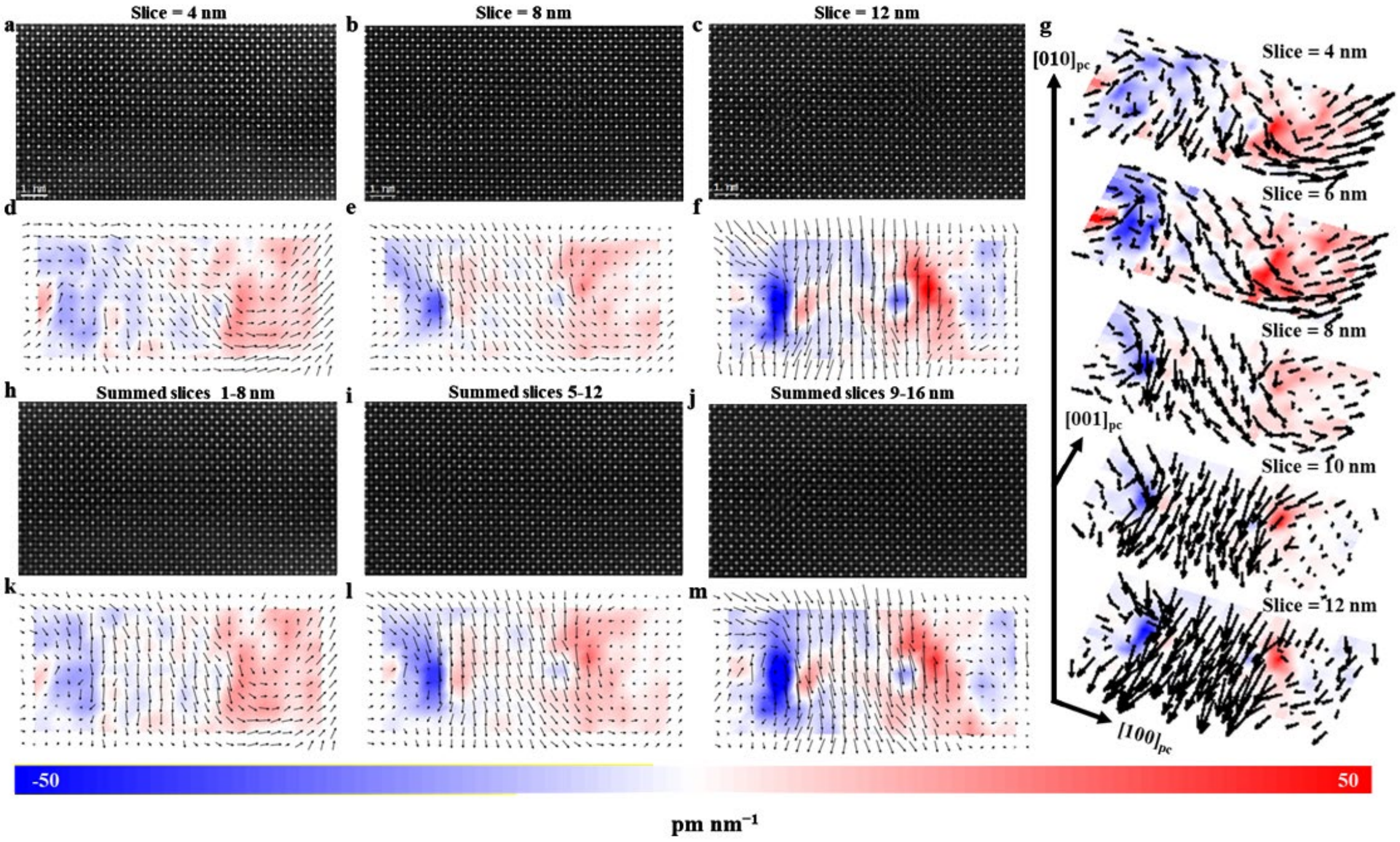

**Figure 4 | Experimental mapping of polarization topology along the depth direction.** (a–c) Reconstructed phase-contrast slices extracted at depths of 4, 8, and 12 nm, respectively. (d–f) Corresponding Pb displacement vector maps relative to the oxygen sublattice, overlaid with the associated vorticity distributions, revealing pronounced depth-dependent variations in the polarization topology. (g) 3D stack of the depth-dependent evolution of the polar pattern. (h–j) Phase-contrast reconstructions integrated over different depth ranges (1–8 nm, 5–12 nm, and 9–16 nm). (k–m) Corresponding Pb displacement vector maps and vorticity distributions, demonstrating how depth integration modifies the apparent polarization topology and can obscure underlying depth-dependent polar configurations.

The depth-summed phase image in Figure 3e exhibits elongated atomic-column contrast, particularly within regions 1 and 2, highlighted by dashed rectangles. Representative slices extracted at depths of 5, 10, and 13 nm reveal distinctly different local atomic configurations. In region 1, no significant lattice distortion is observed at a depth of 5 nm (Figure 3f); however, clear splitting of the Ti columns emerges at 10 nm (Figure 3g) and disappears again at 13 nm (Figure 3h). In contrast, region 2 exhibits relatively little structural variation at 5 nm (Figure 3i), while split Ti and Pb column features become apparent at 10 nm (Figure 3j) and remain visible at 13 nm (Figure 3k). Subtle variations in the O-site contrast are also visible in some of these regions, particularly in Figures 3h and 3j. These observations demonstrate substantial spatial heterogeneity in the axial distribution of atomic displacements and indicate that the underlying lattice distortions evolve significantly as a function of depth. The corresponding *c*-axis projection of the reconstructed volume (Figure 3i) exhibits less pronounced structural variation than that observed along the *a*-axis direction. Nevertheless, subtle depth-dependent changes in atomic-column contrast are clearly resolved within the magnified region shown in Figure 3d, corresponding to the area highlighted by the green dashed rectangle in Figure 3i. These observations indicate that variations associated with the a-axis displacement are present but considerably weaker than those associated with the c-axis displacement, possibly due to the confinement imposed by the adjacent STO layers. Collectively, these results are fully consistent with the depth-sectioning ADF-STEM measurements and provide direct, real-space evidence of pronounced depth-dependent atomic rearrangements within the PTO layer.

The pronounced structural variations observed throughout the reconstructed volume naturally raise the question of how depth-dependent atomic displacements influence the local polarization topology. To address this issue, we quantitatively mapped Ti displacements relative to the surrounding Pb sublattice (Figure S10) and Pb displacements relative to the oxygen sublattice (Figure S11) throughout the entire reconstructed slices (Figure S9). The resulting Ti-displacement maps (Figure S10) reveal a strong depth-dependent polarization landscape. At depths of approximately 3-5 nm, a well-defined vortex-like polarization structure is observed on the right side of the field of view, whereas no comparable feature is present on the right. As the reconstruction depth increases to around 7 nm, the vortex gradually weakens. By approximately 10 nm, the original vortex-like structure has largely disappeared and transformed into a sinusoidal polarization configuration. Remarkably, near the bottom surface of the specimen, a new vortex-like feature emerges on the left side of the field of view, revealing a substantial depth-dependent reorganization of the local polarization topology.

A key advantage of MEP is its ability to simultaneously resolve both heavy cations (Pb and Ti) and light elements (O), thereby providing comprehensive structural information that cannot be readily accessed using conventional STEM imaging modes. To further investigate the origin of the observed polarization evolution, representative slices extracted at depths of approximately 4, 8 and 12 nm were selected for detailed analysis (Figures 4a–c). The corresponding Pb-displacement fields referenced to the oxygen sublattice (Figures 4d–f) exhibit similarly rich depth-dependent behavior. At a depth of approximately 4 nm, a weak vortex-like feature is observed on the left side of the field of view, while the right side is dominated by a sinusoidal polarization configuration. At intermediate depths, 7 nm, the left-side vortex disappears, whereas the right-side flux-closure-like pattern evolves into a more rotational polarization arrangement. At approximately 8 nm, both vortex-like and sinusoidal polarization features become significantly weakened. By 12 nm, the vortex-like feature re-emerges on the left side, while the polarization texture on the right develops into a substantially more complex configuration, resulting in a polarization state markedly different from those observed near either specimen surface. The three-dimensional evolution of these polarization textures is directly visualized in the volumetric rendering shown in Figure 4g. To evaluate the effect of depth integration on the observed polarization patterns, phase-contrast reconstructions were summed over different depth intervals (Figures 4h–j and S12). While the reconstruction integrated over the entire specimen thickness (1–16 nm) still exhibits an apparent vortex-like polarization pattern, integration over narrower depth ranges (1–8 nm, 5–12 nm, and 9–16 nm) yields distinctly different polarization configurations, in excellent agreement with the depth-resolved evolution described above. This comparison further demonstrates that projection through the entire specimen thickness can obscure the pronounced depth dependence of the local polarization texture.

These observations demonstrate that the projection-based measurements can obscure substantial variations in polarization occurring along the beam direction. Polarization structures that appear stable and well defined in conventional projected images may, in reality, be confined to specific depth ranges, evolve significantly throughout the specimen thickness, or even disappear entirely within the interior of the material. The pronounced depth-dependent evolution revealed here provides direct evidence for the inherently three-dimensional nature of polarization topology in ferroelectric superlattices. More broadly, these findings underscore the critical importance of depth-resolved structural characterization for accurately describing emergent polar states and establish MEP as a powerful approach for uncovering hidden three-dimensional polarization textures that remain inaccessible to conventional projection-based techniques.

**Conclusions:**

In summary, we demonstrate that polarization structures in complex oxides exhibit pronounced three-dimensional variations that remain largely inaccessible to conventional projection-based imaging methods. By combining depth-sectioning LAADF STEM, HAADF STEM, and MEP, we directly resolve the depth-dependent evolution of atomic structure in a $[(PTO)_{15}/(STO)_{15}]_{15}$ superlattice and quantitatively map the corresponding polarization topology throughout the reconstructed volume. The three-dimensional MEP reconstruction reveals substantial variations in the displacements of Pb, Ti, and O sublattices along the beam direction, giving rise to significant changes in the local polarization configuration as a function of depth.

Notably, vortex-like polarization features are observed near the specimen surfaces but become strongly suppressed within the interior of the reconstructed volume, where distinct polarization textures emerge. These observations demonstrate that polarization topologies inferred from conventional projected images can originate from the superposition of multiple depth-dependent polarization states and therefore may not faithfully represent the underlying three-dimensional polar structure. The results further reveal that polarization configurations that appear robust in projection can be highly localized in depth and undergo substantial evolution across the specimen thickness.

Our findings provide direct experimental evidence of previously hidden three-dimensional polarization textures and establish depth-resolved electron microscopy as a powerful platform for visualizing and quantifying complex polar topologies atomic-scale resolution. More broadly, this work highlights the importance of three-dimensional structural characterization for understanding the relationship between local atomic distortions and emergent polarization states in ferroic materials. The ability to directly access depth-dependent polarization textures opens new opportunities for uncovering hidden topological states, elucidating structure-property relationships, and guiding the design of next-generation functional oxide heterostructures with topology-driven properties.

## Methods

### Sample

$[(PTO)_{15}/(STO)_{15}]_{15}$ superlattices were epitaxially grown on (001)pc $DyScO_3$ (DSO) substrates by 90º off-axis radiofrequency (RF) magnetron sputtering. The sputtering system was equipped with two independent guns, employing ceramic targets of stoichiometric $SrTiO_3$ and $Pb_{1.2}TiO_3$, the latter containing a 20% excess of Pb to compensate for its volatility during growth. Prior to deposition, the DSO was treated to achieve well-defined -DyO and -$TiO_2$ single-terminated surfaces, respectively, following previously established procedures.

All the samples were deposited at an RF power of 100 W, under a total pressure of 200 mTorr in a mixed $O_2$/Ar atmosphere with a 9:51 ratio. The deposition temperature was set to 625 ºC.

The deposition rates of $PbTiO_3$ and $SrTiO_3$ under such conditions were calibrated by ex-situ X-ray reflectivity, yielding ≈0.018 and 0.012 nm/s, respectively, which enabled precise control of superlattice periodicity. The structural quality and periodicity of the superlattices were confirmed by high-resolution X-ray diffraction and HAADF-STEM[40]

**STEM characterizations**

High-quality cross-sectional specimens were prepared using a Thermo Fisher Helios 5 UX dual-beam focused ion beam (FIB) system. STEM experiments were performed on a double aberration-corrected Thermo Fisher Spectra 300 microscope operated at 300 kV. Depth-sectioning LAADF-STEM and HAADF-STEM were conducted by acquiring a series of atomic-resolution images at different defocus planes, enabling structural characterization at different depths within the specimen. LAADF-STEM images were acquired using a probe convergence semi-angle of 20.5 mrad and an annular detector collection semi-angle of 24–47 mrad. HAADF-STEM images were acquired using a probe convergence semi-angle of 20.6 mrad and an annular detector collection semi-angle of 63–200 mrad. Energy-dispersive X-ray spectroscopy (EDX) and electron energy-loss spectroscopy (EELS) measurements were performed using an FEI Titan G2 80–200 ChemiSTEM microscope. The LAADF-STEM, HAADF-STEM and the 4D-STEM images simulations, were performed using the multislice electron scattering formalism implemented in the abTEM Python library[45]. Part of the data processing and analysis was performed using the py4DSTEM package[46].

**Multislice Electron Ptychography (MEP)**

Four-dimensional (4D) STEM dataset for MEP was acquired on an EMPAD detector. The probe convergence semi-angle was set to 20.5 mrad. The probe was overfocused by approximately 20 nm relative to the top surface of the sample and scanned over a raster grid with a step size of 0.52 Å. Each diffraction frame consisted of 128 × 128 pixels with a reciprocal-space pixel size of 0.043 $Å^{-1}$. Optimal parameters for the ptychographic reconstruction were determined using a Bayesian optimization framework[47] implemented in the *fold-slice* package[48-49]. Ptychographic reconstructions were performed using Ptyrad[50].

**Polarization analysis**

The atomic positions of Pb, Ti, and O columns were determined from the reconstructed phase-contrast images using a custom analysis workflow implemented in Python based on the Atomap package[51]. Local polarization-related displacement fields were subsequently quantified by measuring the relative positions of the atomic columns. The Ti displacement vector was calculated as the deviation of each Ti column from the centroid of the four surrounding Pb columns, whereas the Pb displacement vector was calculated as the deviation of each Pb column from the centroid of the four surrounding O columns. The corresponding vorticity distributions were calculated from the displacement vector fields using a previously reported method[3], with the fitting region extending to the eight surrounding atomic columns.

**ACKNOWLEDGMENT**

The authors gratefully acknowledge Dr. Lei Jin for helpful discussions on the simulations and assistance with the EDX measurements.

**Funding Statement**

We acknowledge ICN2, the Joint Electron Microscopy Center at ALBA (JEMCA) and the Ernst Ruska-Centre for Microscopy and Spectroscopy with Electron at Forschungszentrum Jülich for providing key facilities and technical guidance. ICN2 is supported by the Severo Ochoa program from Spanish MCIN/AEI (Grant No. CEX2021-001214-S), and the CERCA Programme, Generalitat de Catalunya. ICN2 and ER-C is aare founding members of e-DREAM[51]. J.A. and X.H. acknowledge funding from Generalitat de Catalunya (Grant No. 2021SGR00457). This study is part of the Advanced Materials programme and was supported by MCIN with funding from European Union NextGenerationEU (PRTR-C17.I1) and by Generalitat de Catalunya (In-CAEM Project). We acknowledge support from CSIC Interdisciplinary Thematic Platform (PTI+) on Quantum Technologies (PTI-QTEP+). This work has been funded by the European Commission – NextGenerationEU (Regulation EU 2020/2094), through CSIC's Quantum Technologies Platform (QTEP). ICN2 acknowledges funding from Grant IU16-014206 (METCAM-FIB) funded by the European Union through the European Regional Development Fund (ERDF), with the support of the Ministry of Research and Universities, Generalitat de Catalunya. F. R. acknowledges support from Ministerio de Ciencia (Spain), projects PID2022-138883NB-I00, PID2021-128281NA-I00, TED2021-130930B-I00, Xunta de Galicia (Centro de investigación do Sistema universitario de Galicia accreditation 2023-2027, ED431G 2023/03) and the European Union (European Regional Development Fund -ERDF). The research of F. R. receives financial support from the Oportunius Program, Xunta de Galicia. N. V.-D. acknowledges financial support from MINECO (Spain) through an FPI fellowship (PRE2020-096467). CBE acknowledges support for this research through a Vannevar Bush Faculty Fellowship (ONR N00014-20-1-2844), the Gordon and Betty Moore Foundation's EPiQS Initiative, Grant GBMF9065 and the US National Science Foundation (NSF) through the Designing Materials to Revolutionize and Engineer our Future (DMREF) program (NSF award DMREF-2522669). Thin film synthesis at the University of Wisconsin–Madison was supported by the US Department of Energy (DOE), Office of Science, Office of Basic Energy Sciences (BES), under award number DE-FG02-06ER46327. This work is in the framework of the Universitat Autonoma de Barcelona Materials Science PhD program. X.H. acknowledges PhD scholarship support from the China Scholarship Council (CSC) (Grant No. 202304910019), scholarship support from the German Academic Exchange Service (DAAD) (Grant No. 57811724), and support from the Severo Ochoa Centre of Excellence programme (Grant No. CEX2021-001214-S/MICIU/AEI/10.13039/501100011033). Y.L. acknowledges the National Natural Science Foundation of China (Grant Nos. 12402183), the Severo Ochoa Seed Funding program (Grant CEX2021-001214-S/MICIU/AEI/10.13039/501100011033). The authors acknowledge the use of the Spectra 300 microscope, provided under ALBA Synchrotron proposal (Grant No. 20240320035, 20250340231 and 20260400042). GC acknowledges support from the EU through HORIZON-MSCA-SE-3D-TOPO action (Grant agreement ID: 101236483) and from Spain's MICINN through grant PID2023-148673NB-I00 (Project GRIPHO2)

**Author contributions**

Experimental measurements were performed at ICN2, the Joint Electron Microscopy Center at ALBA (JEMCA) and the Ernst Ruska-Centre for Microscopy and Spectroscopy with Electron at Forschungszentrum Jülich. All institutions contributed substantially to the development and completion of this work. X.H. conceived the original idea for this work and designed the project under the supervision of J.A., G.C., and R.D-B. X.H., J.A., P.H., conducted STEM experiments. X.H. performed data analysis on the MEP reconstruction, simulation and polarization mapping with the suggestions from P.H.,Y.L. and Z.S. N.V A. E. C. B-E and F. R.. fabricated the samples. F.R., and J.S. provided assistance with the experiments. P.L help with the ptychography reconstruction and simulation. X.H. wrote the manuscript with the help from all authors. All authors contributed to the discussion of the results and the revision of the manuscript.

**Competing interests**

The authors declare that they have no competing interests.

**Data and materials availability**

All data are available in the manuscript or the supplementary information.

## REFERENCES

1. Gradauskaite, E.; Meier, Q. N.; Gray, N.; Sarott, M. F.; Scharsach, T.; Campanini, M.; Moran, T.; Vogel, A.; Del Cid-Ledezma, K.; Huey, B. D.; Rossell, M. D.; Fiebig, M.; Trassin, M., Defeating depolarizing fields with artificial flux closure in ultrathin ferroelectrics. *Nat Mater* **2023,** *22*, 1492-1498.
2. Tang, Y. L.; Zhu, Y. L.; Ma, X. L.; Borisevich, A. Y.; Morozovska, A. N.; Eliseev, E. A.; Wang, W. Y.; Wang, Y. J.; Xu, Y. B.; Zhang, Z. D.; Pennycook, S. J., Observation of a periodic array of flux-closure quadrants in strained ferroelectric $PbTiO_3$ films. **2015,** *348*, 547-551.
3. Yadav, A. K.; Nelson, C. T.; Hsu, S. L.; Hong, Z.; Clarkson, J. D.; Schleputz, C. M.; Damodaran, A. R.; Shafer, P.; Arenholz, E.; Dedon, L. R.; Chen, D.; Vishwanath, A.;

Minor, A. M.; Chen, L. Q.; Scott, J. F.; Martin, L. W.; Ramesh, R., Observation of polar vortices in oxide superlattices. *Nature* **2016,** *530*, 198-201.
4. Han, L.; Addiego, C.; Prokhorenko, S.; Wang, M.; Fu, H.; Nahas, Y.; Yan, X.; Cai, S.; Wei, T.; Fang, Y.; Liu, H.; Ji, D.; Guo, W.; Gu, Z.; Yang, Y.; Wang, P.; Bellaiche, L.; Chen, Y.; Wu, D.; Nie, Y.; Pan, X., High-density switchable skyrmion-like polar nanodomains integrated on silicon. *Nature* **2022,** *603*, 63-67.
5. Das, S.; Tang, Y. L.; Hong, Z.; Goncalves, M. A. P.; McCarter, M. R.; Klewe, C.; Nguyen, K. X.; Gomez-Ortiz, F.; Shafer, P.; Arenholz, E.; Stoica, V. A.; Hsu, S. L.; Wang, B.; Ophus, C.; Liu, J. F.; Nelson, C. T.; Saremi, S.; Prasad, B.; Mei, A. B.; Schlom, D. G.; Iniguez, J.; Garcia-Fernandez, P.; Muller, D. A.; Chen, L. Q.; Junquera, J.; Martin, L. W.; Ramesh, R., Observation of room-temperature polar skyrmions. *Nature* **2019,** *568*, 368-372.
6. Nahas, Y.; Prokhorenko, S.; Fischer, J.; Xu, B.; Carretero, C.; Prosandeev, S.; Bibes, M.; Fusil, S.; Dkhil, B.; Garcia, V.; Bellaiche, L., Inverse transition of labyrinthine domain patterns in ferroelectric thin films. *Nature* **2020,** *577*, 47-51.
7. Eliseev, E. A.; Fomichov, Y. M.; Kalinin, S. V.; Vysochanskii, Y. M.; Maksymovich, P.; Morozovska, A. N., Labyrinthine domains in ferroelectric nanoparticles: Manifestation of a gradient-induced morphological transition. *Physical Review B* **2018,** *98,* 054101.
8. Wang, Y. J.; Feng, Y. P.; Zhu, Y. L.; Tang, Y. L.; Yang, L. X.; Zou, M. J.; Geng, W. R.; Han, M. J.; Guo, X. W.; Wu, B.; Ma, X. L., Polar meron lattice in strained oxide ferroelectrics. *Nat Mater* **2020,** *19*, 881-886.
9. Luk'yanchuk, I., Tikhonov, Y., Razumnaya, A., Vinokur, V. M., Hopfions emerge in ferroelectrics. *Nat Commun* **2020**, 11, 2433.
10. Wei, X. K.; Jia, C. L.; Sluka, T.; Wang, B. X.; Ye, Z. G.; Setter, N., Neel-like domain walls in ferroelectric $Pb(Zr,Ti)O_3$ single crystals. *Nat Commun* **2016,** *7*, 12385.
11. Prokhorenko, S.; Nahas, Y.; Bellaiche, L., Fluctuations and Topological Defects in Proper Ferroelectric Crystals. *Phys Rev Lett* **2017,** *118*, 147601.
12. Peters, J. J. P.; Apachitei, G.; Beanland, R.; Alexe, M.; Sanchez, A. M., Polarization curling and flux closures in multiferroic tunnel junctions. *Nat Commun* **2016,** *7*, 13484.
13. Rogers, A.; Holsgrove, K.; Schafer, N. A.; Koppitz, B.; McCluskey, C. J.; Yedama, S.; Lynch, R.; Sloan, K.; Porter, B.; Sykes, A.; Catalan, A.; Silva, R. S., Jr.; Bruno, F. Y.; Seddon, S. D.; Lu, H.; Ruesing, M.; Fink, C.; Fahler-Muenzer, P.; Fearn, S.; Heutz, S. E. M.; Hadjimichael, M.; Ramasse, Q. M.; Alexe, M.; Kumar, A.; McQuaid, R. G. P.; Gruverman, A.; Sanna, S.; Eng, L. M.; Gregg, J. M., Polar discontinuities, emergent conductivity, and critical twist-angle-dependent behaviour at wafer-bonded ferroelectric interfaces. *Nat Commun* **2026,** *17*, 1842.
14. Zubko, P.; Wojdel, J. C.; Hadjimichael, M.; Fernandez-Pena, S.; Sene, A.; Luk'yanchuk, I.; Triscone, J. M.; Iniguez, J., Negative capacitance in multidomain ferroelectric superlattices. *Nature* **2016,** *534*, 524-8.
15. Xu, R.; Karthik, J.; Damodaran, A. R.; Martin, L. W., Stationary domain wall contribution to enhanced ferroelectric susceptibility. *Nat Commun* **2014,** *5*, 3120.
16. Stachiotti, M. G.; Sepliarsky, M., Toroidal ferroelectricity in $PbTiO_3$ nanoparticles. *Phys Rev Lett* **2011,** *106*, 137601.
17. Liu, Y.; Zhang, H.; Shapovalov, K.; Niu, R.; Cairney, J. M.; Liao, X.; Roleder, K.; Majchrowski, A.; Arbiol, J.; Ghosez, P.; Catalan, G., Vortices and antivortices in antiferroelectric PbZrO(3). *Nat Mater* **2025,** *24*, 1359-1363.

18. Jeong, C., Lee, J., Jo, H., Oh, J., Baik, H., Go, K., Son, J., Choi, Y., Prosandeev, S., Bellaiche, L., Yang. Y., Revealing the three-dimensional arrangement of polar topology in nanoparticles. *Nat Commun* **2024,** 15, 3887.
19. Marti-Sanchez, S.; Botifoll, M.; Oksenberg, E.; Koch, C.; Borja, C.; Spadaro, M. C.; Di Giulio, V.; Ramasse, Q.; Garcia de Abajo, F. J.; Joselevich, E.; Arbiol, J., Sub-nanometer mapping of strain-induced band structure variations in planar nanowire core-shell heterostructures. *Nat Commun* **2022,** *13*, 4089.
20. Liu, Y.; Niu, R.; Majchrowski, A.; Roleder, K.; Cordero-Edwards, K.; Cairney, J. M.; Arbiol, J.; Catalan, G., Translational Boundaries as Incipient Ferrielectric Domains in Antiferroelectric $PbZrO_3$. *Phys Rev Lett* **2023,** *130*, 216801.
21. Li, X.; Chen, S.; Li, M.; Liu, K.; Bai, X.; Gao, P., Atomic origin of Ti-deficient dislocation in $SrTiO_3$ bicrystals and their electronic structures. *Journal of Applied Physics* **2019,** *126,* 174106.
22. Gao, P.; Ishikawa, R.; Feng, B.; Kumamoto, A.; Shibata, N.; Ikuhara, Y., Atomic-scale structure relaxation, chemistry and charge distribution of dislocation cores in $SrTiO_3$. *Ultramicroscopy* **2018,** *184*, 217-224.
23. Ishikawa, R.; Shibata, N.; Taniguchi, T.; Ikuhara, Y., Three-Dimensional Imaging of a Single Dopant in a Crystal. *Physical Review Applied* **2020,** *13*, 034064.
24. Sanchez-Santolino, G.; Rouco, V.; Puebla, S.; Aramberri, H.; Zamora, V.; Cabero, M.; Cuellar, F. A.; Munuera, C.; Mompean, F.; Garcia-Hernandez, M.; Castellanos-Gomez, A.; Iniguez, J.; Leon, C.; Santamaria, J., A 2D ferroelectric vortex pattern in twisted $BaTiO_3$ freestanding layers. *Nature* **2024,** *626*, 529-534.
25. Lun, Y. Z.; Hu, X. X.; Ren, Q.; Saeed, U.; Gupta, K.; Mundet, B.; Pinto-Huguet, I.; Santiso, J.; Padilla-Pantoja, J.; Manuel Caicedo Roque, J.;, Ma, Y. P.; Li, Q.; Tang, G.; Pesquera, D.; Wang, X. Y.; Hong, J.; Arbiol, J.; Catalan, G.; Polarization Vortices in a Ferromagnetic Metal via Twistronics. **2025**, arXiv:2505.17742
26. Borisevich, A. Y.; Lupini, A. R.; Pennycook, S. J., Depth sectioning with the aberration-corrected scanning transmission electron microscope. **2006,** *103*, 3044-3048.
27. Kp, H.; Wei, X.; Lee, C. H.; Yoon, D.; Lee, Y.; Crust, K. J.; Shao, Y. T.; Xu, R.; Kang, J. H.; Liang, C.; Park, J.; Hwang, H. Y.; Muller, D. A., Mind the Gap-Imaging Buried Interfaces in Twisted Oxide Moires. *Adv Mater* **2026,** *38*, e21189.
28. Chen, Z.; Jiang, Y.; Shao, Y.-T.; Holtz, M. E.; Odstrčil, M.; Guizar-Sicairos, M.; Hanke, I.; Ganschow, S.; Schlom, D. G.; Muller, D. A., Electron ptychography achieves atomic-resolution limits set by lattice vibrations. **2021,** *372*, 826-831.
29. Sha, H.; Zhang, Y.; Ma, Y.; Li, W.; Yang, W.; Cui, J.; Li, Q.; Huang, H.; Yu, R., Polar vortex hidden in twisted bilayers of paraelectric $SrTiO_3$. *Nat Commun* **2024,** *15*, 10915.
30. Dong, Z.; Huo, M.; Li, J.; Li, J.; Li, P.; Sun, H.; Gu, L.; Lu, Y.; Wang, M.; Wang, Y.; Chen, Z., Visualization of oxygen vacancies and self-doped ligand holes in $La_3Ni_2O_{7-\delta}$. *Nature* **2024,** *630*, 847-852.
31. Ribet, S. M.; Varnavides, G.; Pedroso, C. C. S.; Cohen, B. E.; Ercius, P.; Scott, M. C.; Ophus, C., Uncovering the three-dimensional structure of upconverting core–shell nanoparticles with multislice electron ptychography. *Applied Physics Letters* **2024,** *124*, 240601.

32. Zhang, Y.; Ahammed, B.; Bae, S. H.; Lee, C.-H.; Huang, J.; Hossain, M. A.; Rakib, T.; van der Zande, A. M.; Ertekin, E.; Huang, P. Y., Atom-by-atom imaging of moiré phasons with electron ptychography. **2025,** *389*, 423-428.
33. Chen, Z.; Shao, Y.-T.; Jiang, Y.; Muller, D., Three-dimensional imaging of single dopants inside crystals using multislice electron ptychography. *Microscopy and Microanalysis* **2021,** *27*, 2146-2148.
34. Xu, C.; Luo, N.; Yue, J.; Chen, C.; Bian, T.; Zhang, C.; Che, X.; Liang, J.; Li, M. M.; Yin, J.; Chen, Z.; Zhang, S.; Pan, X.; Zhu, Y., Intrinsic polar vortex crystals in A-site layer-ordered perovskites. *Nature* **2026,** *653*, 83-89.
35. Kp, H.; Xu, R.; Patel, K.; Crust, K. J.; Khandelwal, A.; Zhang, C.; Prosandeev, S.; Zhou, H.; Shao, Y. T.; Bellaiche, L.; Hwang, H. Y.; Muller, D. A., Electron ptychography reveals a ferroelectricity dominated by anion displacements. *Nat Mater* **2025,** *24*, 1433-1440.
36. Zhong, H.; Wang, S.; Zhang, Q. H.; Liu, Z. H.; Xie, D. G.; Lu, J. L.; Jin, S.; Zhang, S.; Guo, E. J.; He, M.; Wang, C.; Gu, L.; Yang, G. Z.; Jin, K. J.; Ge, C. Observation of One-Dimensional Charged Domain Walls in Ferroelectric $ZrO_2$. *Science* **2026**, *391*, 407–411.
37. Stoica, V.A., Laanait, N., Dai, C., Yuan, Z., Zhang, Z., Lei, S., McCarter, M. R., Yadav, A., Damodaran, A. R., Das, S., Stone, G. A., Karapetrova, J., Walko, D. A. , Zhang, X., Martin, L. W., Ramesh, R., Chen, L.-Q., Wen, H., Gopalan, V., Freeland, J. W., Optical creation of a supercrystal with three-dimensional nanoscale periodicity. *Nat. Mater.* **2019,** 18, 377–383 .
38. Phillips, P. J.; De Graef, M.; Kovarik, L.; Agrawal, A.; Windl, W.; Mills, M. J., Atomic-resolution defect contrast in low angle annular dark-field STEM. *Ultramicroscopy* **2012,** *116*, 47-55.
39. Grillo, V.; Rossi, F., A new insight on crystalline strain and defect features by STEM–ADF imaging. *Journal of Crystal Growth* **2011**, *318* (1), 1151-1156.
40. Varela-Domínguez, N., Claro, M., Gutiérrez-Llorente, A., Langenberg, E., Hu, X. X., Bagués, N., Edgeton, A., Beom-Eom, C., Arbiol, J., Santiso, J., Rivadulla, F., Complex polar superstructure controlled thermal conductivity in ferroelectric PbTiO3/SrTiO3 superlattices, arXiv, doi.org/10.48550/arXiv.2607.08683
41. Yamashita, S.; Kikkawa, J.; Yanagisawa, K.; Nagai, T.; Ishizuka, K.; Kimoto, K., Atomic number dependence of Z contrast in scanning transmission electron microscopy. *Sci Rep* **2018,** *8*, 12325.
42. Nellist, P.; Pennycook, S.; The principles and interpretation of annular dark-field Z-contrast imaging, *Adv. Imag. Elec. Phys* **2000***,* 113, 147-203
43. Shibata, N.; Findlay, S. D.; Kohno, Y.; Sawada, H.; Kondo, Y.; Ikuhara, Y., Differential phase-contrast microscopy at atomic resolution. *Nature Physics* **2012,** *8*, 611-615.
44. Okunishi, E.; Ishikawa, I.; Sawada, H.; Hosokawa, F.; Hori, M.; Kondo, Y., Visualization of Light Elements at Ultrahigh Resolution by STEM Annular Bright Field Microscopy. *Microscopy and Microanalysis* **2009,** *15*, 164-165.
45. Madsen, J.; Susi, T., The abTEM code: transmission electron microscopy from first principles. *Open Res Eur* **2021,** *1*, 24.
46. Savitzky, B. H.; Zeltmann, S. E.; Hughes, L. A.; Brown, H. G.; Zhao, S.; Pelz, P. M.; Pekin, T. C.; Barnard, E. S.; Donohue, J.; Rangel DaCosta, L.; Kennedy, E.; Xie, Y.; Janish, M. T.; Schneider, M. M.; Herring, P.; Gopal, C.; Anapolsky, A.; Dhall, R.;

Bustillo, K. C.; Ercius, P.; Scott, M. C.; Ciston, J.; Minor, A. M.; Ophus, C., py4DSTEM: A Software Package for Four-Dimensional Scanning Transmission Electron Microscopy Data Analysis. *Microsc Microanal* **2021,** *27*, 712-743.
47. Zhang, C.; Shao, Y.-T.; Baraissov, Z.; Duncan, C. J.; Hanuka, A.; Edelen, A. L.; Maxson, J. M.; Muller, D. A., Bayesian Optimization for Multi-dimensional Alignment: Tuning Aberration Correctors and Ptychographic Reconstructions. *Microscopy and Microanalysis* **2022,** *28*, 3146-3148.
48. Wakonig, K.; Stadler, H. C.; Odstrcil, M.; Tsai, E. H. R.; Diaz, A.; Holler, M.; Usov, I.; Raabe, J.; Menzel, A.; Guizar-Sicairos, M., PtychoShelves, a versatile high-level framework for high-performance analysis of ptychographic data. *J Appl Crystallogr* **2020,** *53*, 574-586.
49. Thibault, P.; Menzel, A., Reconstructing state mixtures from diffraction measurements. *Nature* **2013,** *494* (7435), 68-71.
50. Lee, C. H.; Zeltmann, S. E.; Yoon, D.; Ma, D.; Muller, D. A., PtyRAD: A High-Performance and Flexible Ptychographic Reconstruction Framework with Automatic Differentiation. *Microsc Microanal* **2025,** *31*, ozaf070.
51. Nord, M.; Vullum, P. E.; MacLaren, I.; Tybell, T.; Holmestad, R., Atomap: a new software tool for the automated analysis of atomic resolution images using two-dimensional Gaussian fitting. *Adv Struct Chem Imaging* **2017,** *3*, 9.
52. Ciancio, R.; Dunin-Borkowski, R. E.; Snoeck, E.; Kociak, M.; Holmestad, R.; Verbeeck, J.; Kirkland, A. I.; Kothleitner, G.; Arbiol, J., e-DREAM: the European Distributed Research Infrastructure for Advanced Electron Microscopy. *Microscopy and Microanalysis* **2022,** *28*, 2900-2902.

# Supplementary information for

# Depth-Resolved Evolution of Buried Polar Topologies in a $PbTiO_3/SrTiO_3$ Superlattice

Xinxin Hu[1,*], Penghan Lu[2], Noa Varela-Dominguez[3], Anthony Edgeton[4], Chang Beom-Eom[4], Francisco Rivadulla[3], José Santiso[1], Yingzhuo Lun[1,6], Zhihua Sun[5], Rafal Dunin-Borkowski[2], Gustau Catalan[1,7], Jordi Arbiol[1,7,*]

[1]Catalan Institute of Nanoscience and Nanotechnology - ICN2 (CSIC & BIST), Barcelona, Catalonia 08193, Spain

[2]Ernst Ruska-Centre for Microscopy and Spectroscopy with Electrons, Forschungszentrum Jülich, Jülich, Germany

[3] CiQUS, Centro Singular de Investigacion en Quimica Bioloxica e Materiais Moleculares, Departamento de Quimica-Fisica, Universidade de Santiago de Compostela, Santiago de Compostela 15782, Spain

[4]Department of Materials Science and Engineering, University of Wisconsin-Madison, Madison, WI 53706, USA

[5]Chinese Academy of Sciences Fujian Institute of Research on the Structure of Matter, Fuzhou 350108, China

[6] School of Aerospace Engineering, Beijing Institute of Technology, Beijing 10081, China

[7]Institució Catalana de Recerca i Estudis Avançats (ICREA), Barcelona 08010, Catalonia

*Corresponding authors email: xinxin.hu@icn2.cat, arbiol@icrea.cat

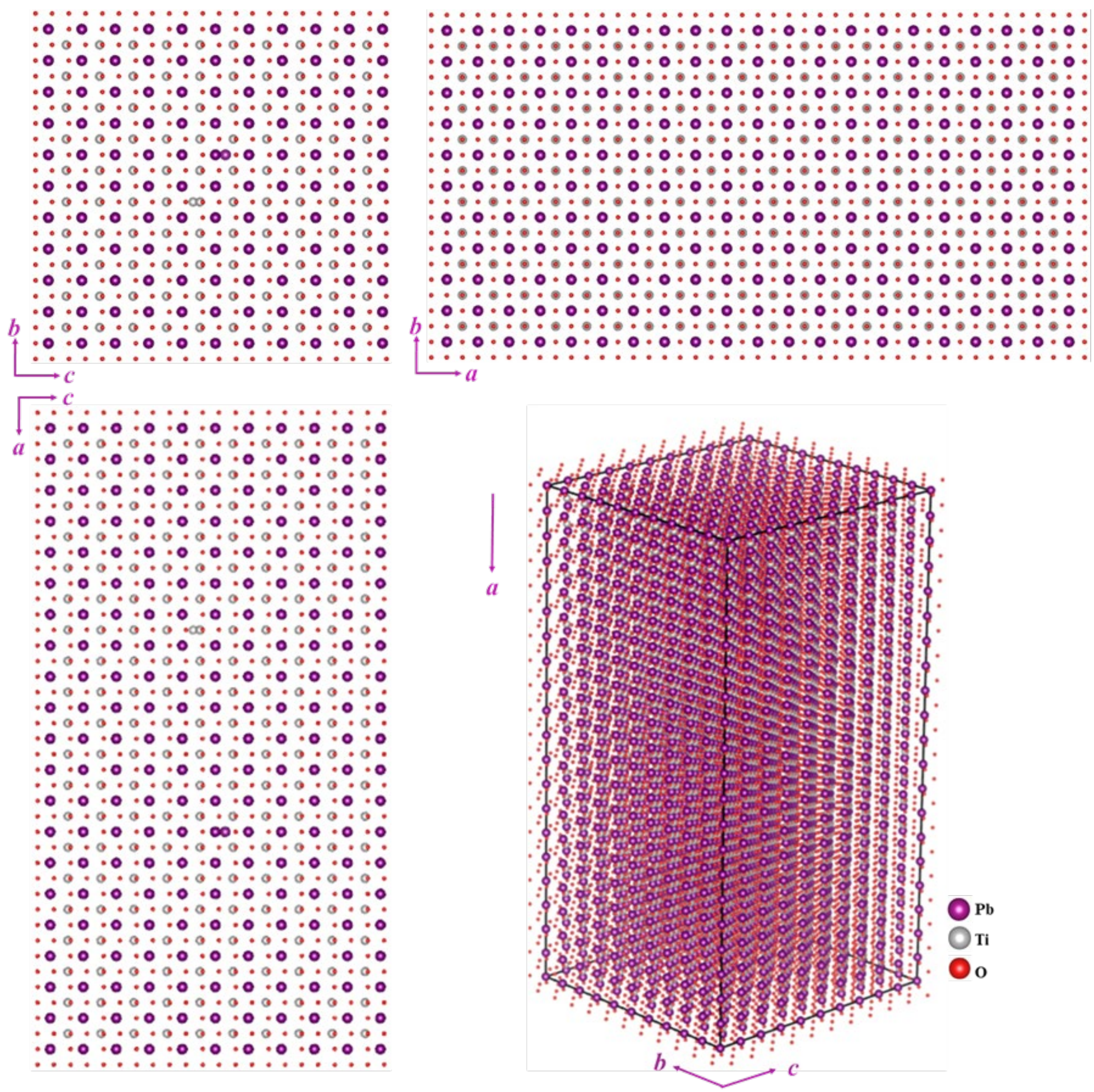

**Figure S1.** PTO model used for LAADF-STEM, HAADF-STEM and MEP simulation viewed along different directions. Dash red and blue circles indicate the shifted Ti and Pb atoms, respectively.

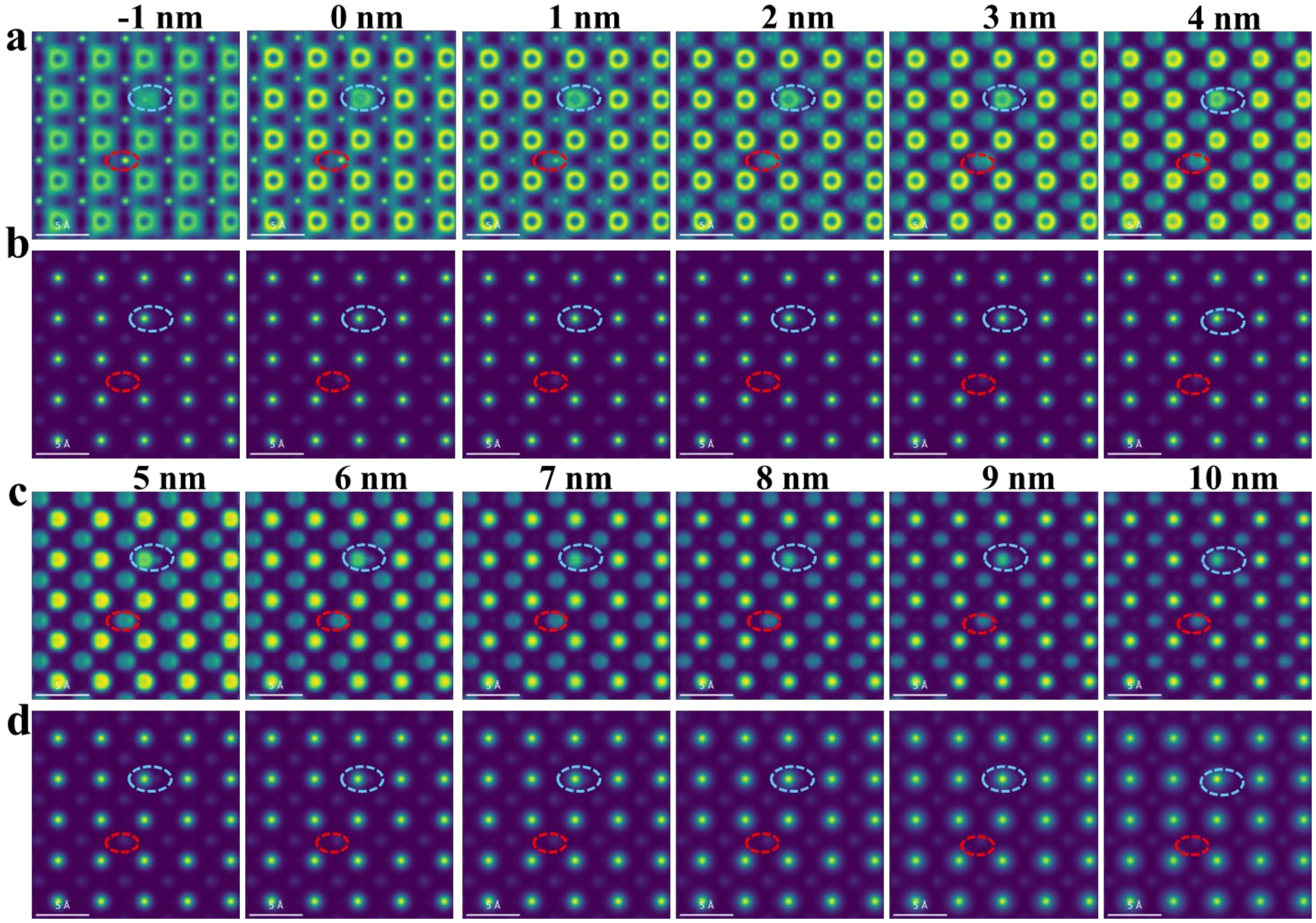

**Figure S2.** Simulated depth-sectioning LAADF-STEM (a, c) and HAADF-STEM (b, d) images of the structural model at different probe focal depths. Dash red and blue circles indicate the position of the shifted Ti and Pb atoms, respectively.

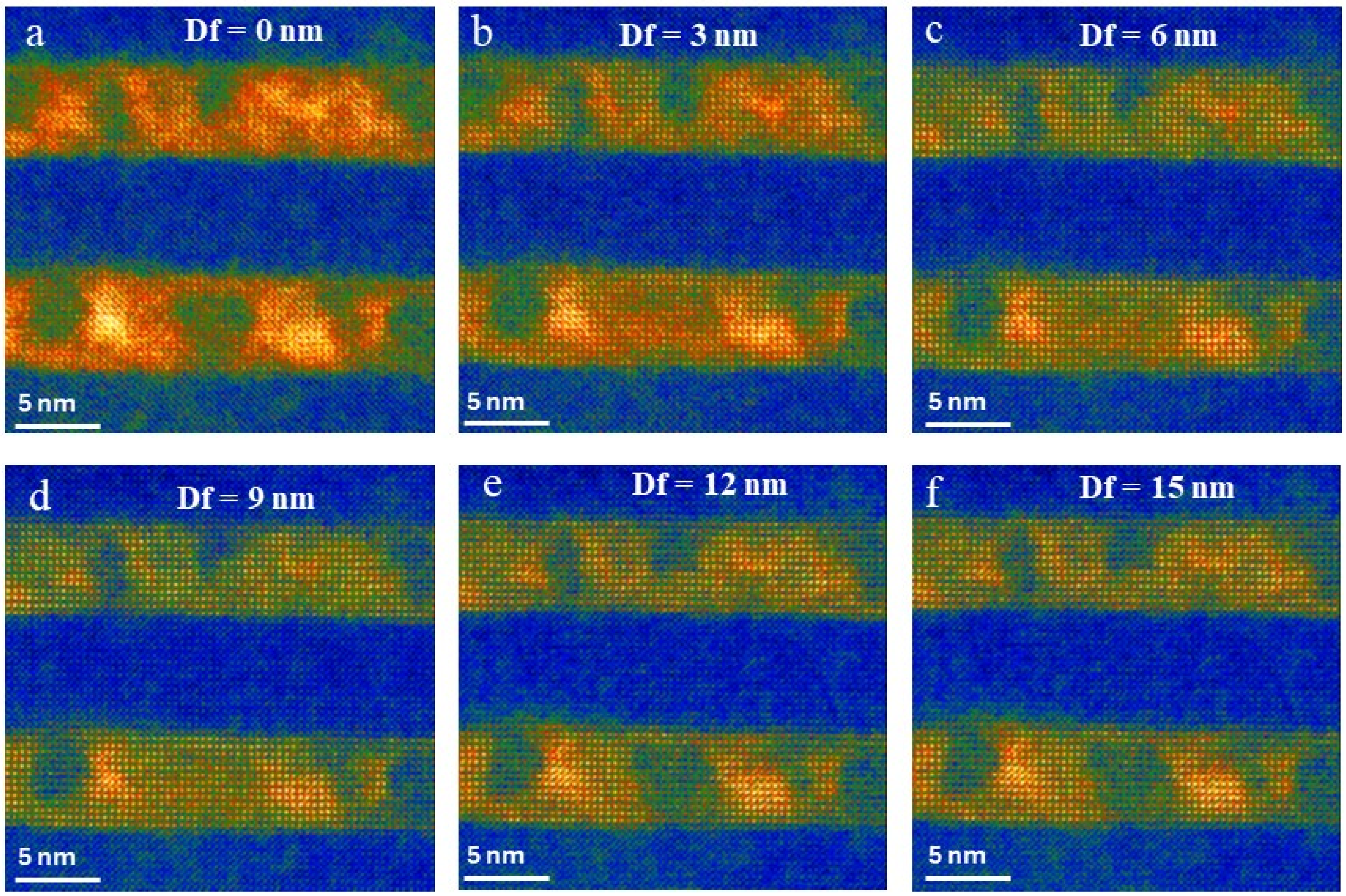

**Figure S3.** LAADF-STEM with different defocus values showing clear contrast variation along depth direction. The bluish contrast is associated with the STO layers, while the orangish contrast is associated with the PTO layers

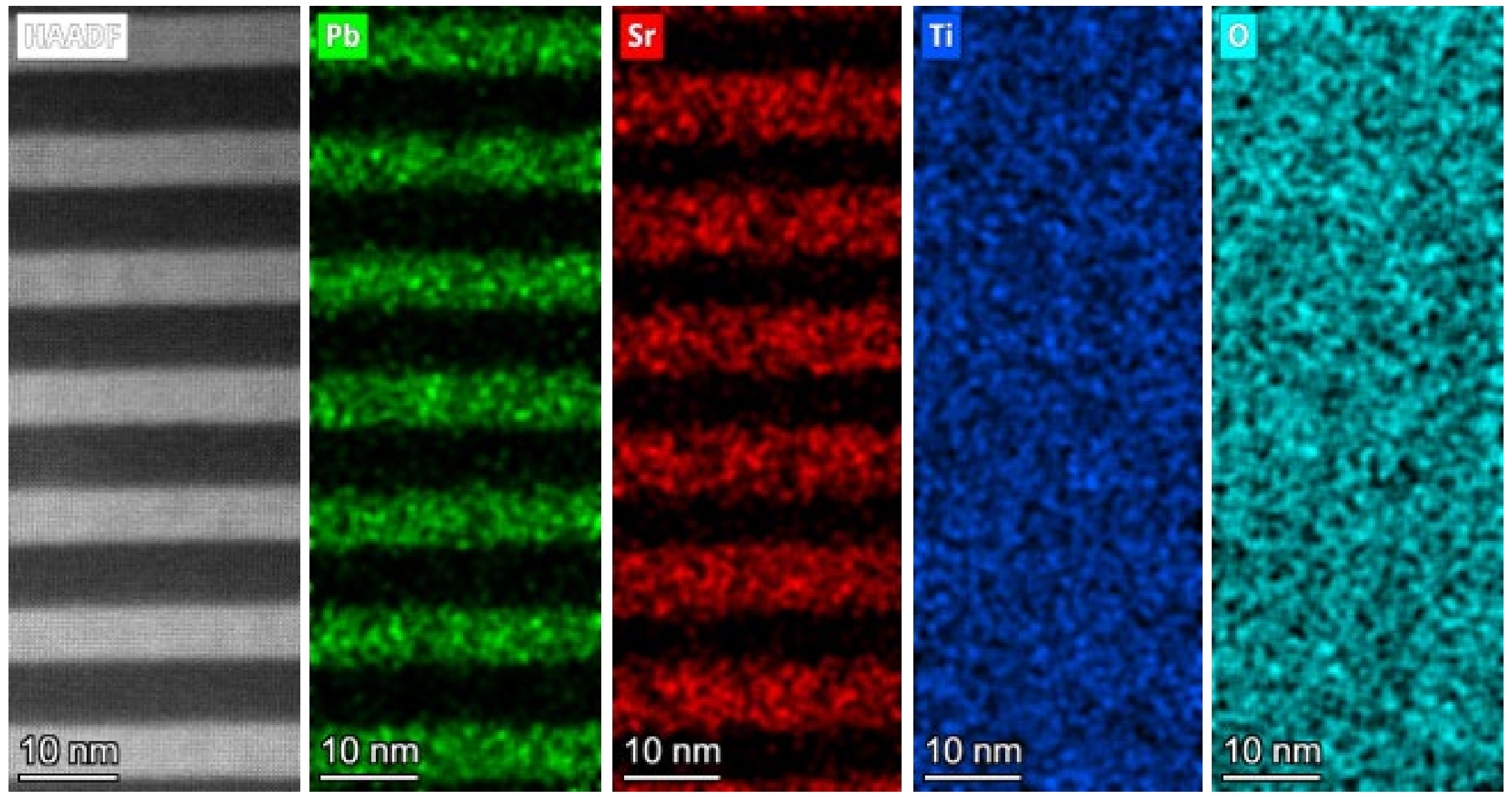

**Figure S4.** HAADF STEM image and corresponding EDX elemental composition maps for Pb, Sr, Ti and O of the PTO/STO superlattice

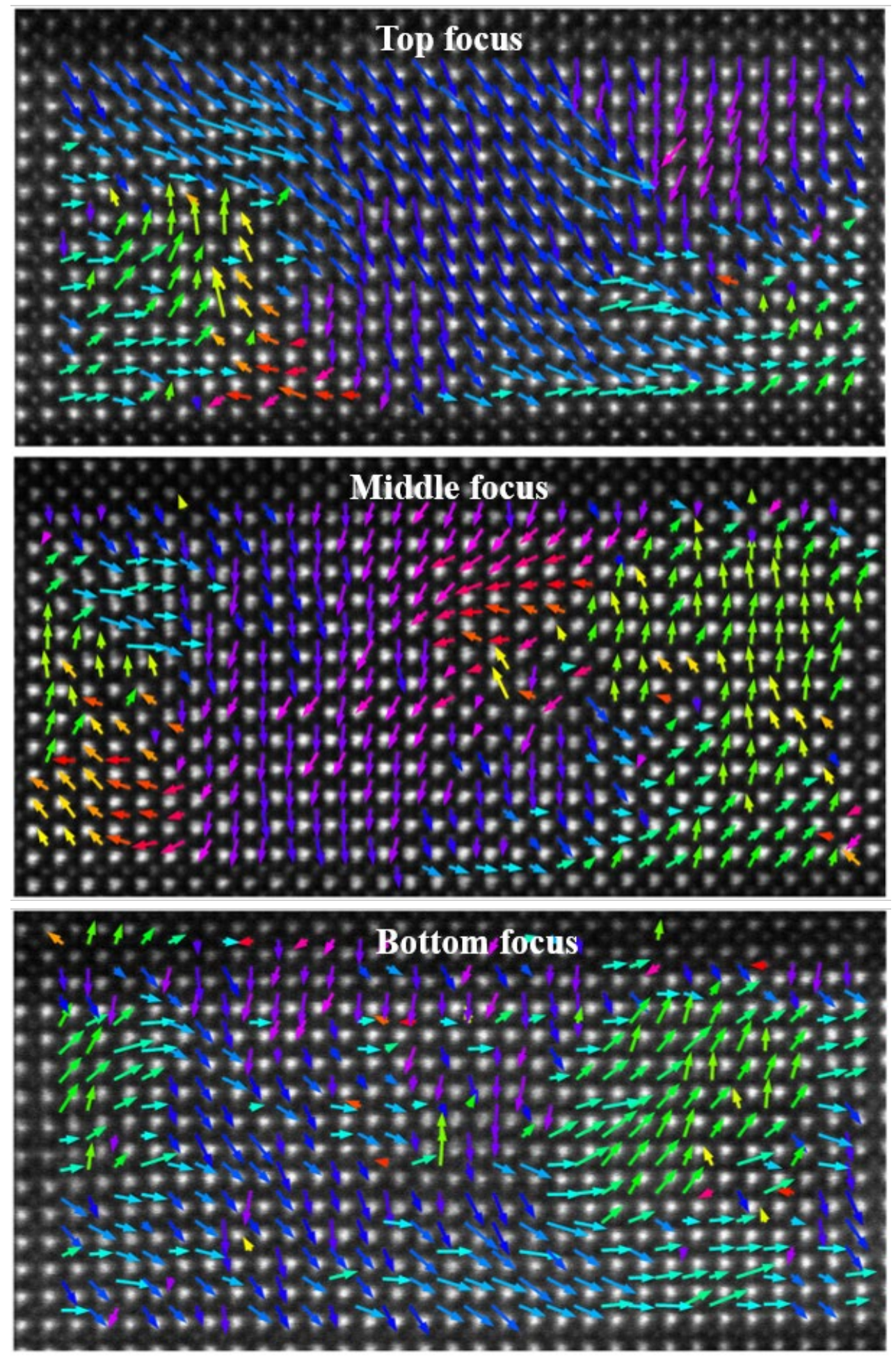

**Figure S5.** Experimental mapping of the depth-resolved polarization topology from HAADF-STEM images.

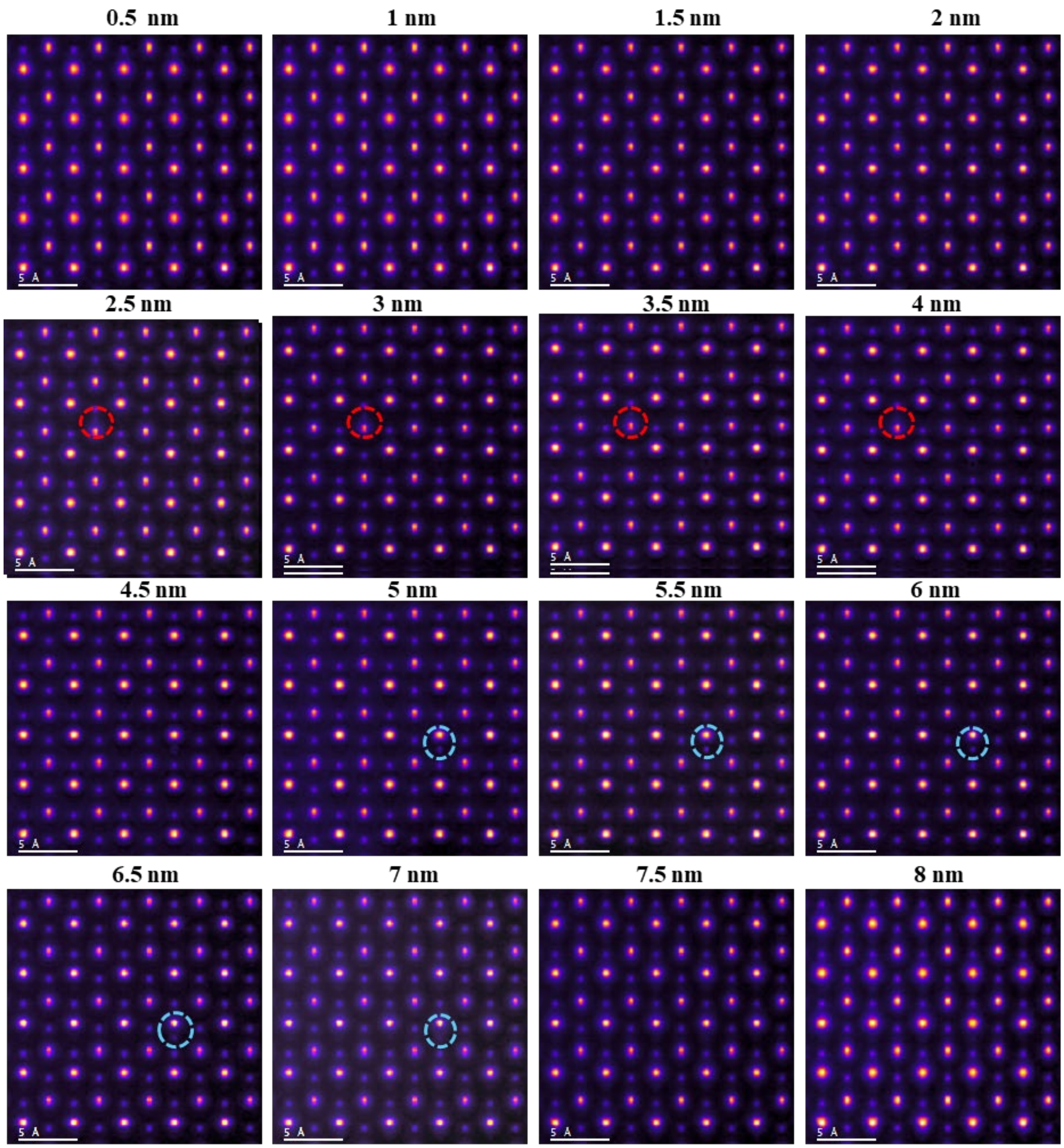

**Figure S6**. MEP reconstructed phase contrast images of the simulated model at different depths. Dash red and blue circles indicate the position of the shifted Ti and Pb atoms, respectively.

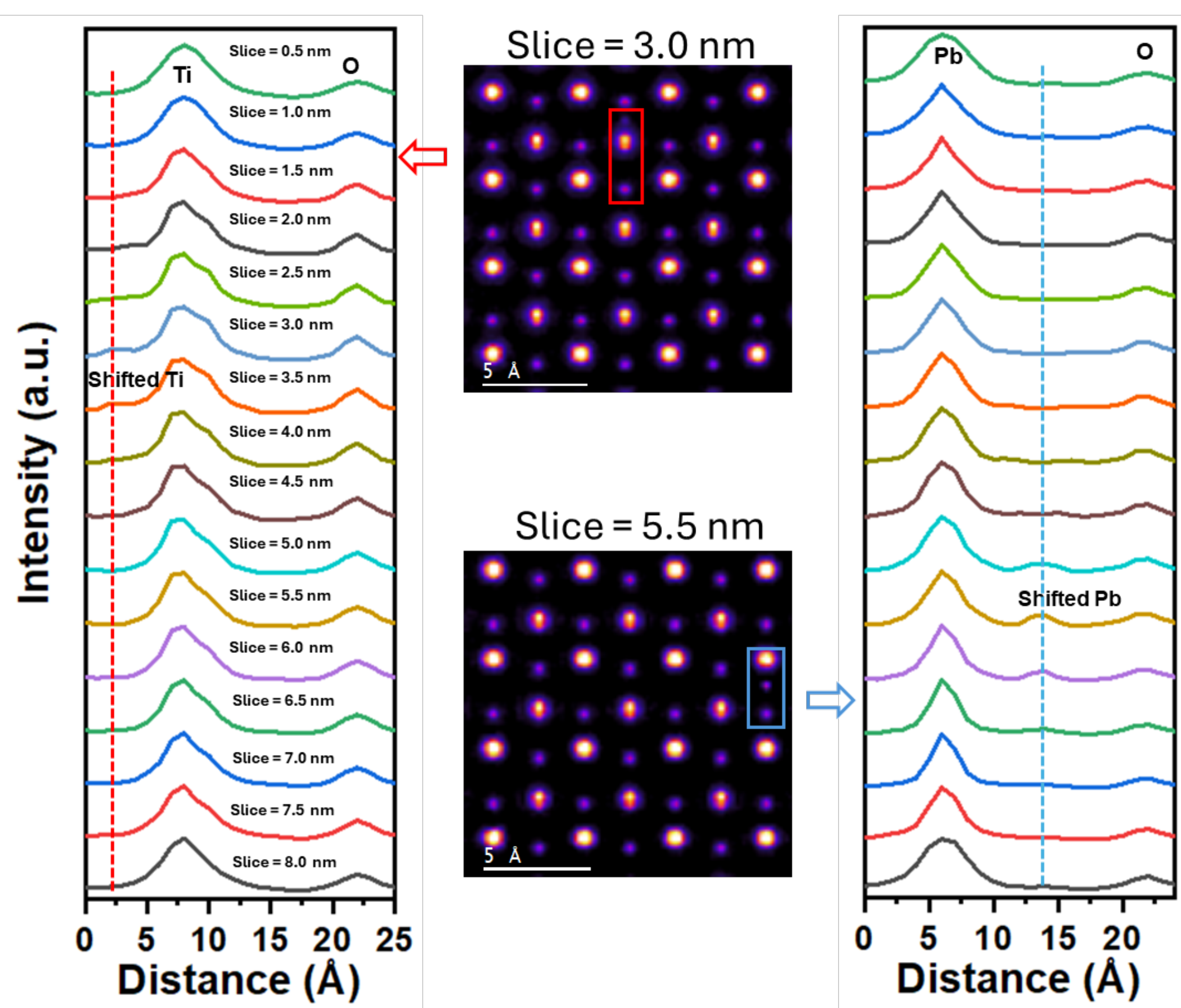

**Figure S7** Linear intensity profiles extracted along the thickness direction from the red and blue rectangular regions, respectively.

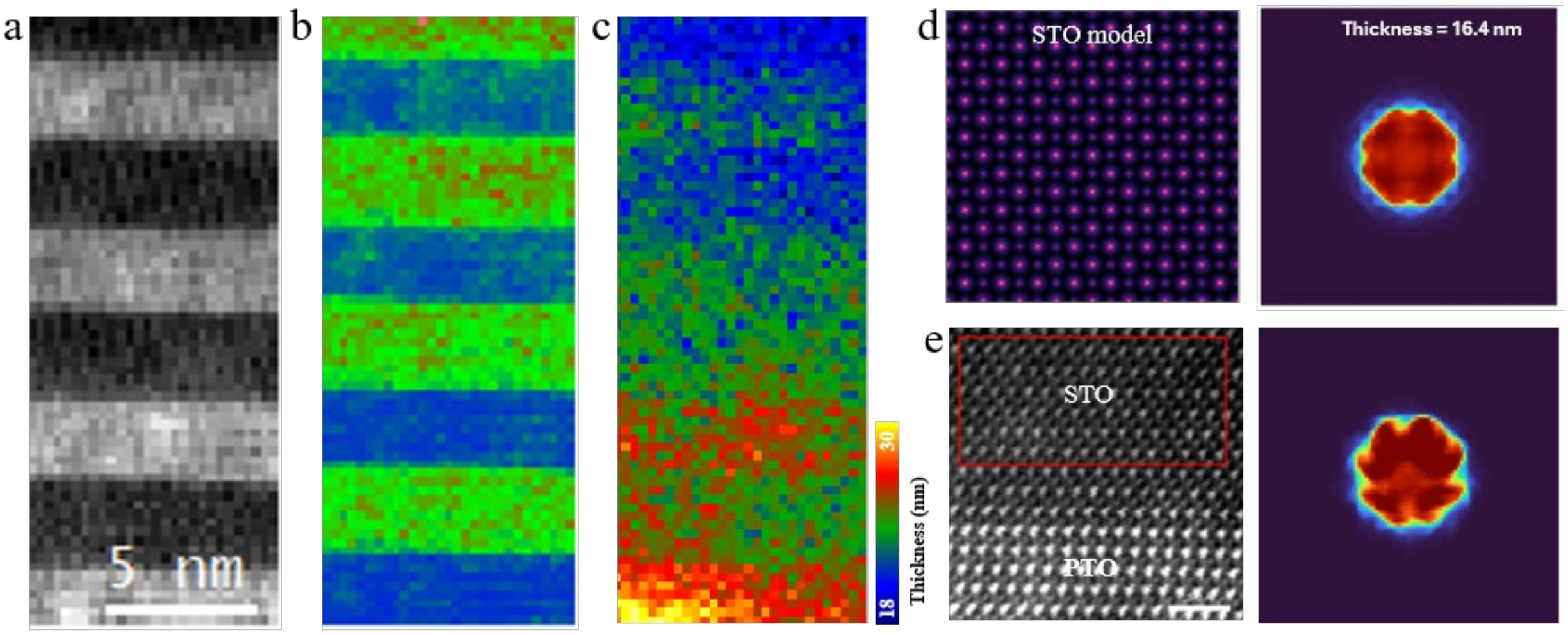

**Figure S8. a**. ADF-STEM **b**. EELS low loss signal intensity and **c,** thickness mapping of the same region from the sample. **d**. simulated PACBED from a 16.4 nm thick STO model and **e.** selected PACBED of STO region from experimental 4DSTEM dataset.

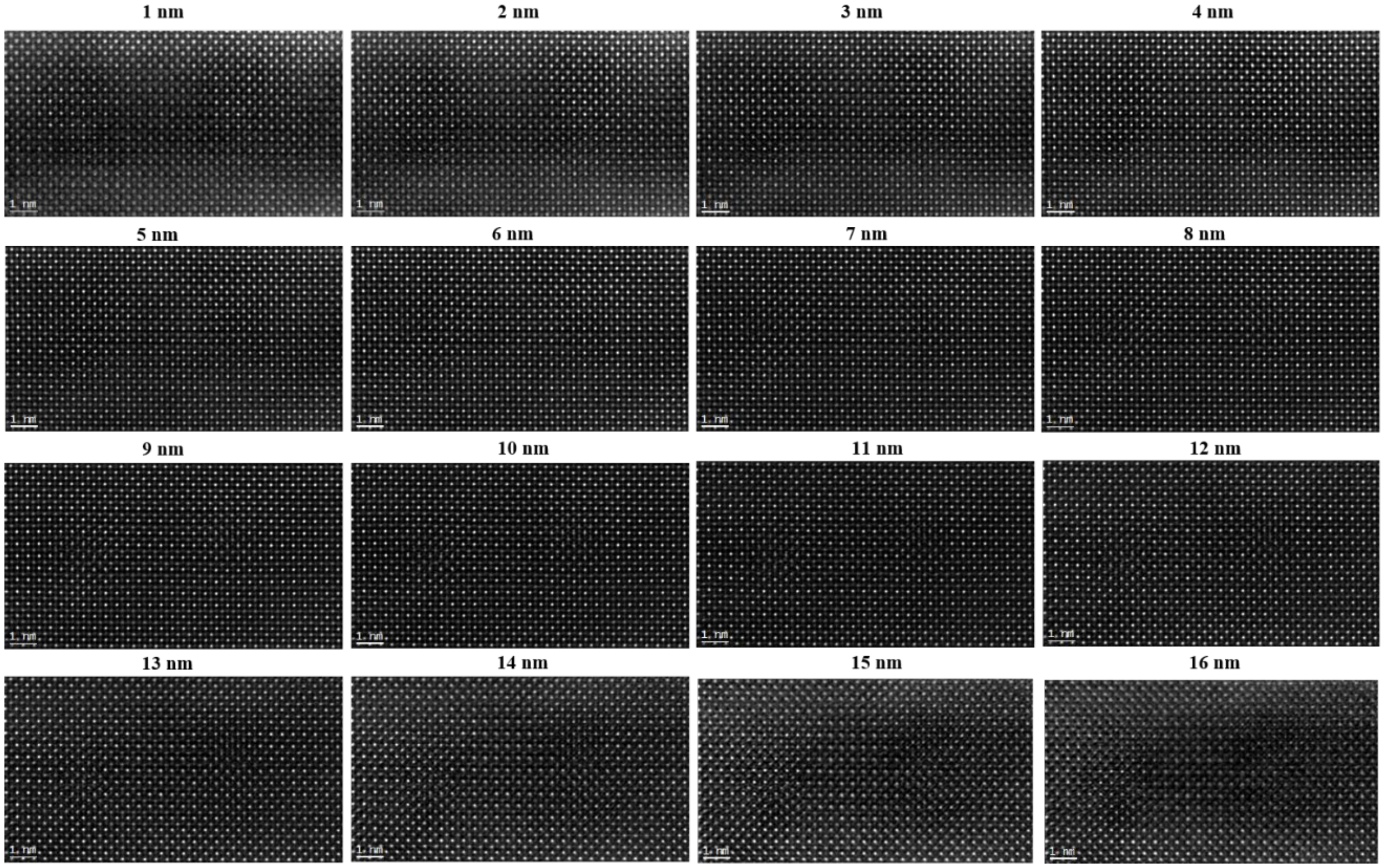

**Figure S9**. Reconstructed phase contrast images of the experimental dataset at different depths.

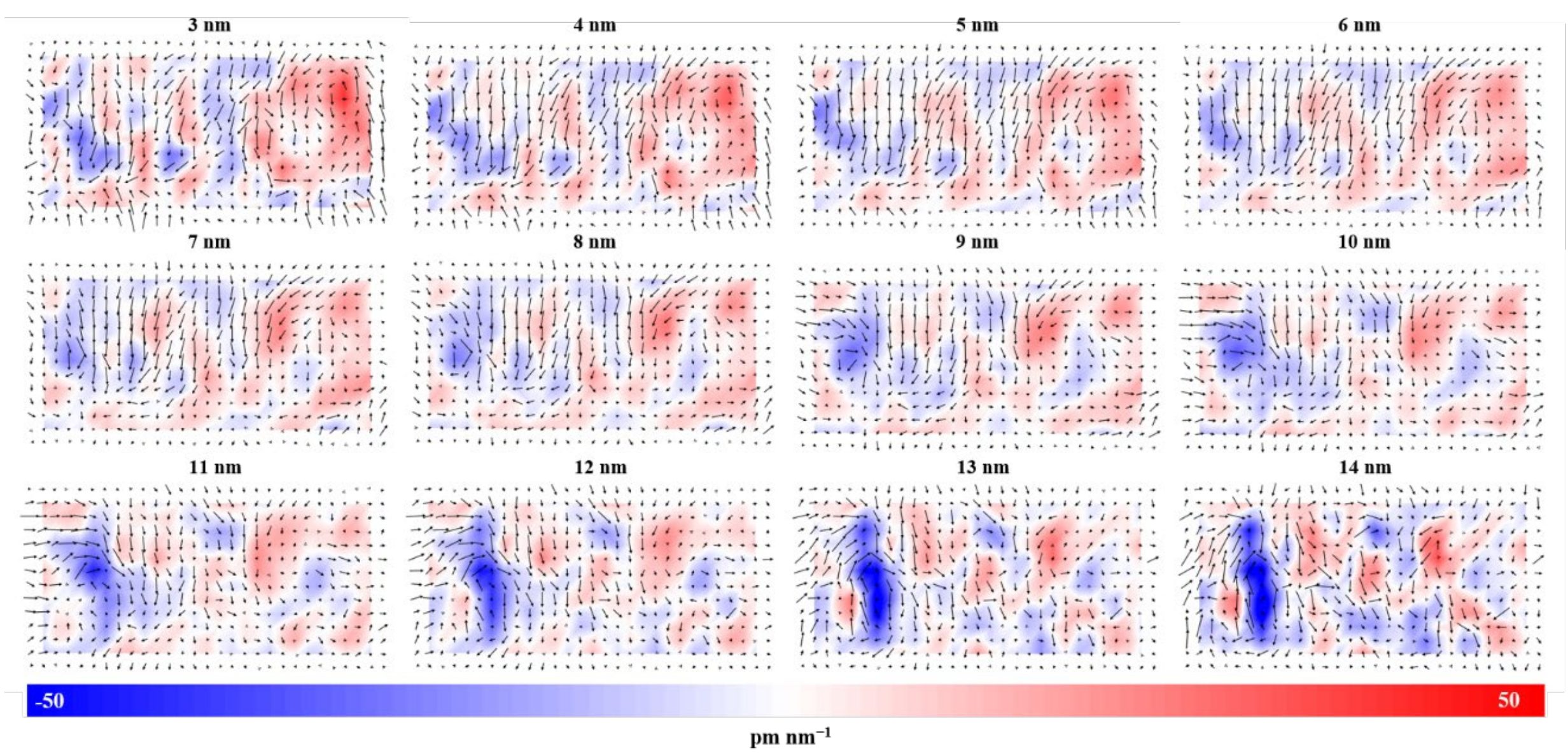

**Figure S10**. Maps of Ti atomic displacements relative to the Pb sublattice from Figure S5, overlaid with the corresponding vorticity distribution.

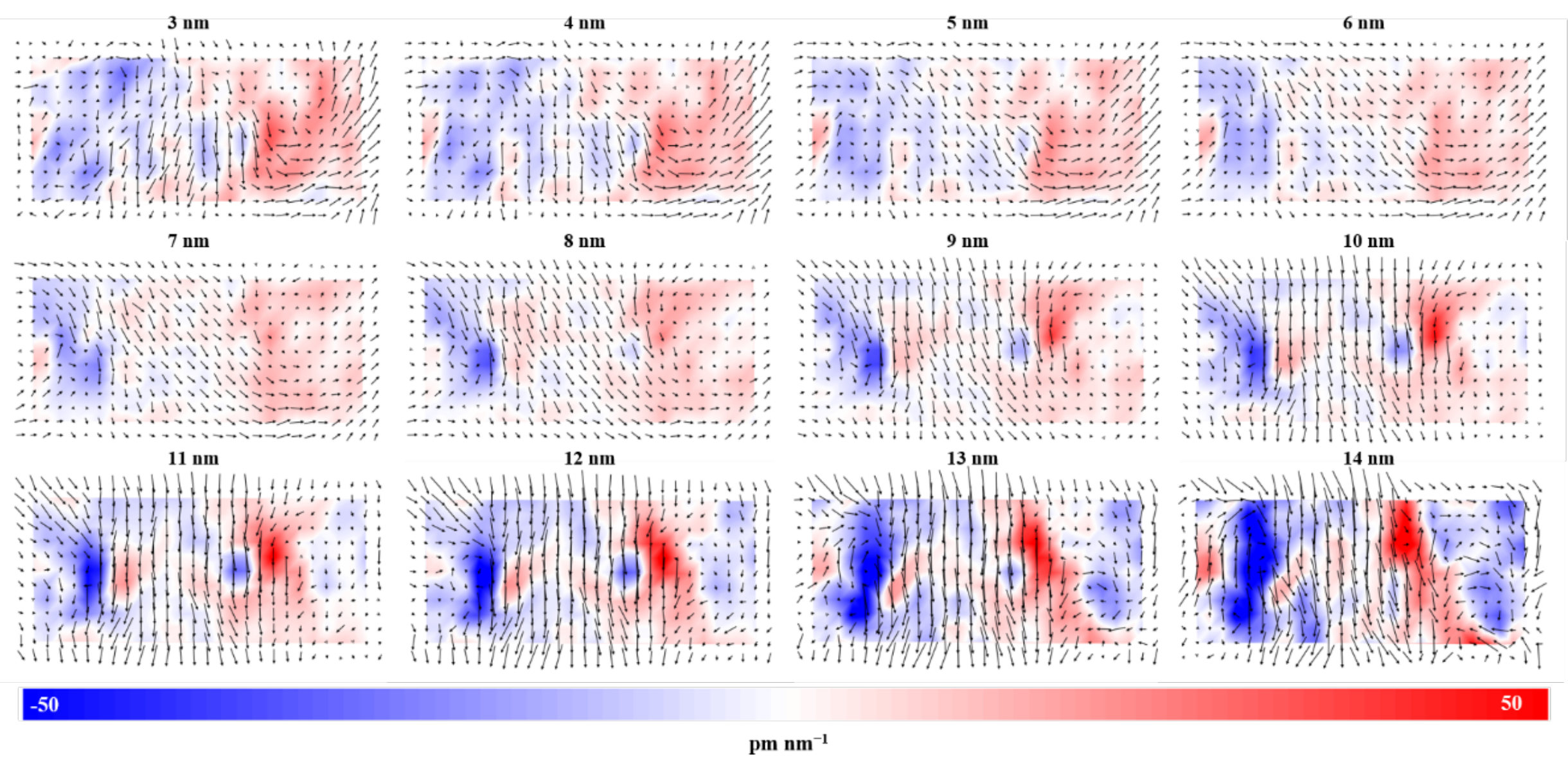

**Figure S11.** Maps of Pb atomic displacements relative to the oxygen sublattice derived from the reconstructed slices shown in Figure S5, overlaid with the corresponding vorticity distributions.

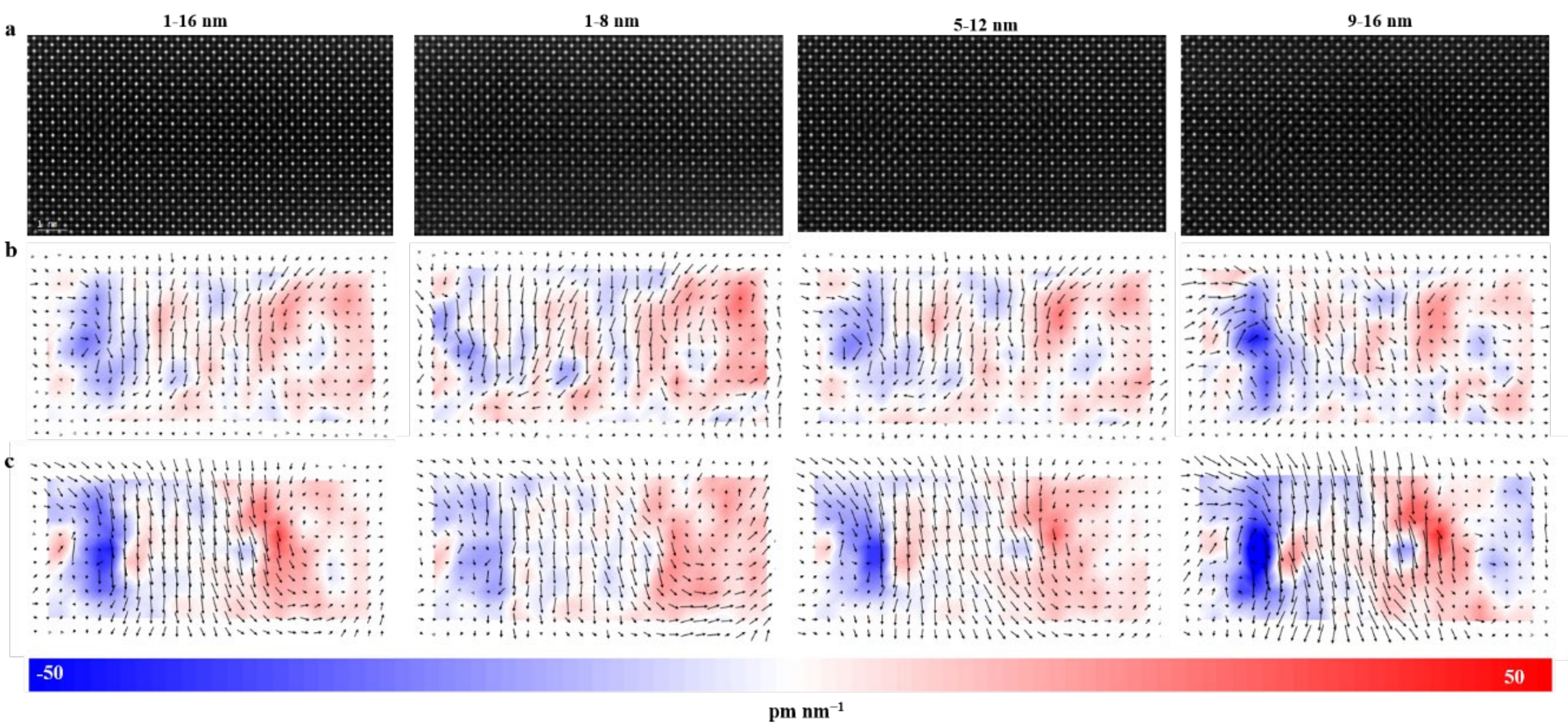

**Figure S12.** (a) Summed phase-contrast images obtained by integrating reconstructed slices over different depth ranges (1–16 nm, 1–8 nm, 5–12 nm, and 9–16 nm). (b) Corresponding Ti displacement vector maps relative to the Pb sublattice and (c) Pb displacement vector maps relative to the oxygen sublattice, together with the associated vorticity distributions. These maps illustrate the influence of depth integration on the observed polarization topology.